\documentclass{article}

\usepackage{amsmath,amsthm}
\usepackage{amssymb}
\usepackage{enumerate}

\usepackage{slashed,enumerate}
\usepackage{hyperref}

\def\titlerunning#1{\gdef\titrun{#1}}
\makeatletter
\def\author#1{\gdef\autrun{\def\and{\unskip, }#1}\gdef\@author{#1}}
\def\address#1{{\def\and{\\\hspace*{18pt}}\renewcommand{\thefootnote}{}%
\footnote {#1}}%
\markboth{\autrun}{\titrun}}
\makeatother
\def\email#1{E-mail: #1}
\def\subjclass#1{\par\medskip
\noindent\textbf{Mathematics Subject Classification (2010).} #1}
\def\keywords#1{\par\medskip
\noindent\textbf{Keywords.} #1}

\numberwithin{equation}{section}

\newcommand{\eq}[1]{\eqref{#1}}
\newcommand{\qtx}[1]{\quad\text{#1}\quad}

\newcommand{\beq}{\begin{equation}}
\newcommand{\eeq}{\end{equation}}

\newcommand{\set}[1]{\left\{ #1 \right\}}

\newcommand{\bl}{\begin{lemma}}
\newcommand{\el}{\end{lemma}}
\newcommand{\bc}{\begin{coro}}
\newcommand{\ec}{\end{coro}}
\newcommand{\bp}{\begin{prop}}
\newcommand{\ep}{\end{prop}}
\newcommand{\bd}{\begin{defini}}
\newcommand{\ed}{\end{defini}}

\DeclareMathOperator{\im}{\mathrm{Im}}  
  
\DeclareMathOperator{\Tr}{\mathrm{Tr}}

\newcommand{\la}{\label}
\newcommand{\ci}{\cite}

\newtheorem{theorem}{Theorem}[section]
\newtheorem{lemma}[theorem]{Lemma}
\newtheorem{coro}[theorem]{Corollary}
\newtheorem{prop}[theorem]{Proposition}
\newtheorem{remark}[theorem]{Remark} 
\newtheorem{defini}[theorem]{Definition}

\newtheorem*{assumptions}{Assumptions}

\newcommand{\De}{\Delta}

\newcommand{\lb}{\lambda}

\newcommand{\bvp}{{\boldsymbol{\varphi}}}

\newcommand{\BPhi}{\mathbf{\Phi}}
\newcommand{\BPsi}{\boldsymbol{\Psi}}
\newcommand{\wBPhi}{\widehat{\BPhi}}
\newcommand{\wbvp}{\widehat{\bvp}}

\newcommand{\La}{\Lambda}

\newcommand{\vpsi}{\overrightarrow{\Psi}}
\newcommand{\vbpsi}{\overrightarrow{\overline{\Psi}}}

\newcommand{\e}{{\rm e}}

\newcommand{\I}{\mathcal{I}}
\newcommand{\Hh}{\mathcal{H}}
\newcommand{\Kk}{\mathcal{K}}

\newcommand{\Bb}{\mathcal{B}}

\newcommand{\Vv}{\mathcal{V}}
\newcommand{\Ww}{\mathcal{W}}
\newcommand{\Ll}{\mathcal{L}}

\newcommand{\Pp}{\mathcal{P}}

\newcommand{\Ss}{\mathcal{S}}
\newcommand{\Ff}{\mathcal{F}}

\newcommand{\RR}{\mathbb{R}}
\newcommand{\NN}{\mathbb{N}}
\newcommand{\ZZ}{\mathbb{Z}}

\newcommand{\CC}{\mathbb{C}}

\newcommand{\PP}{\mathfrak{P}}

\newcommand{\B}{\mathbb  B}

\newcommand{\E}{\mathbb E}
\newcommand{\G}{\mathbb G}
\newcommand{\aaa}{\boldsymbol{a}}
\newcommand{\bbb}{\boldsymbol{b}}
\newcommand{\ccc}{\boldsymbol{c}}

\newcommand{\Sym}{{\rm Sym}}

\newcommand{\sgn}{{\rm sgn}}

\newcommand{\hn}{|\!|\!|}

\newcommand{\hnn}{|\!|\!|\!|}

\newcommand{\od}{\odot}
\newcommand{\ot}{{\od 2}}

\newcommand{\bpart}{\boldsymbol{\partial}}
\newcommand{\bslp}{\boldsymbol{\slashed{\partial}}}

\newcommand{\T}{\mathbb{T}}
\newcommand{\DD}{\mathbb{D}}
\newcommand{\HH}{\mathbb{H}}
\newcommand{\KK}{\mathbb{K}}
\newcommand{\IP}{\mathbb{P}}
\newcommand{\wt}{\widehat{T}}
\newcommand{\wT}{\widehat{\T}}

\newcommand{\f}{\mathbf{f}}
\newcommand{\g}{\mathbf{g}}
\newcommand{\bF}{\mathbf{F}}
\newcommand{\dlt}{\boldsymbol{\delta}}
\newcommand{\bxi}{\boldsymbol{\xi}}
\newcommand{\bth}{\boldsymbol{\theta}}

\newcommand{\one}{\mathbf{1}}

\newcommand{\III}{\boldsymbol{\I}}

\newcommand{\rc}{{\rm c}}
\newcommand{\brc}{\mathbf{c}}

\newcommand{\rrangle}{{\rangle\mkern-4mu\rangle}}
\newcommand{\llangle}{{\langle\mkern-4mu\langle}}

\newcommand{\lt}{\boldsymbol{[\![}}
\newcommand{\rt}{\boldsymbol{]\!]}}

\begin{document}

\titlerunning{Ballistic Behavior on the Bethe Strip}
\title{Ballistic Behavior for Random Schr\"odinger Operators on the Bethe Strip}









\author{Abel Klein\thanks{Supported in part by the NSF under grant DMS-1001509.}
\and 
Christian Sadel}

\date{}

\maketitle

\address{A. Klein:
University of California, Irvine,
Department of Mathematics,
Irvine, CA 92697-3875,  USA.
 \email{aklein@uci.edu}
\and \!\!C. Sadel:
University of California, Irvine,
Department of Mathematics,
Irvine, CA 92697-3875,  USA.
\email{csadel@math.uci.edu}}

\begin{abstract}
The Bethe Strip of width $m$ is the cartesian product $\B\times\{1,\ldots,m\}$, where $\B$ is the Bethe lattice  (Cayley tree).
 We consider  Anderson-like Hamiltonians    $\;H_\lb=\frac12 \De \otimes 1 + 1 \otimes A\,+\,\lb \Vv$ on a 
Bethe strip with connectivity $K \ge 2$, where $A$ is an $m\times m$ symmetric matrix,  $\Vv$ is a random matrix potential, 
and $\lambda$ is the disorder parameter. Under certain conditions on $A$ and $K$, for which we previously 
proved the existence of absolutely continuous spectrum for small $\lb$,
 we now obtain ballistic behavior for the spreading of wave packets evolving under $H_\lb$ for small $\lb$.\\[.2cm]
\end{abstract}

 \subjclass{Primary 82B44; Secondary 47B80, 60H25.}
\keywords{Random Schr\"odinger operators, Anderson model, spreading of wave packets, ballistic behavior, Bethe strip.}


\section{Introduction}


The Bethe lattice (or Cayley tree) $\B$ is an infinite connected graph with no closed loops where each vertex has 
$K+1$ neighbors. $K \in \NN$ is called the connectivity of $\B$. 
The Bethe strip of width $m$ is the cartesian product $\B \times \I$, where 
 $\I=\{1,\ldots,m\}$.
The distance between two sites $x$ and $y$ {of $\B$},  denoted by $d(x,y)$,
is equal to the length of the shortest path connecting $x$ and $y$ in $\B$.
The $\ell^2$ space of functions on the Bethe strip, $\ell^2(\B\times\I)$, can
be identified with the tensor product
$\ell^2(\B)\otimes\CC^m$,  with the direct sum $\bigoplus_{x\in\B} \CC^m$, and with  
$\ell^2(\B,\CC^m)=\big\{u:\B \mapsto \CC^m\,; \sum_{x\in\B} \|u(x)\|^2 < \infty \big\}$,
the space of  $\CC^m$-valued $\ell^2$ functions on $\B$,   {i.e.},
\begin{equation}
\ell^2(\B\times \I)\;\cong\;\ell^2(\B)\otimes\CC^m\;\cong\;\bigoplus_{x\in\B} \CC^m\;\cong\; \ell^2(\B,\CC^m) \;.
\end{equation}

As in \cite{KS} we consider the family of random Hamiltonians on $\ell^2(\B\times\I)$ given by
\begin{equation}\label{Hlambda}
H_\lb\; = \tfrac12\;\De\otimes 1\,+\,1\otimes A\,+\,\lb\Vv \;.
\end{equation}
Here $\De$ denotes the centered Laplacian on $\ell^2(\B)$, which has 
spectrum $ \sigma(\Delta)  =  [-2\sqrt{ K},2\sqrt{K}]\,$ (e.g., \ci{AK}).  We use $\frac{1}{2}\De$
in the definition of $H_\lb$ to simplify some formulas.
$A \in \Sym(m)$ denotes the ``free vertical operator'' on the Bethe strip, where $\Sym(m)\cong \RR^{\frac 1 2 {m(m+1)}}$ is the set of
real symmetric $m\times m$ matrices.
$\Vv$ is the random matrix-potential given by $\Vv=\bigoplus_{x\in\B} V(x)$ on $\bigoplus_{x\in\B} \CC^m$, where 
$\{V(x)\}_{x\in\B}$ are independent identically distributed $\Sym(m)$-valued random variables with common probability distribution $\mu$.
The coefficient  $\lb$ is a real parameter called the {\em disorder}.
In particular, for  $u\in\ell^2(\B,\CC^m)$ we have 
\begin{equation} \label{Hlambda2}
(H_\lb u)(x) = \tfrac 1 2   \!\!\!\!\!
\sum_{\substack{y\in \B\\  d(x,y)=1}}  \!\!\!\!{u(y)}\;+\;A\,u(x)\,+\,\lambda\, V(x)\, u(x)\quad\text{for all}\quad x \in \B\;.
\end{equation}

An important special case of this model is the Anderson model on the product graph $\B\times\G$, where
$\G$ is a finite graph with $m$ labeled vertices.  If $ A_\G$ is the adjacency matrix of the graph $\G$, i.e.,
$(A_\G)_{k,\ell}$ denotes the number of edges between $k\in\G$ and $\ell \in \G$, then
$\De\otimes 1\,+\,1\otimes A_\G$ is the {adjacency} operator on the {product graph} $\B \times \G$. If in \eqref{Hlambda}
  we take
$A=\frac12 A_\G$ and $\mu$   supported 
by the diagonal matrices, with the diagonal entries being independent identically distributed,  then $H_\lb$
is the Anderson model on the product graph $\B\times\G$.  Another special case is the Wegner $m$-orbital model on
the Bethe lattice, obtained by setting $A=0$ and letting  $\mu$  be the probability distribution  of the Gaussian Orthogonal Ensemble (GOE).  This model was  introduced by Wegner \cite{Weg} on the lattice   $\ZZ^d$, where he studied   the limit $m\to\infty$.

There is a  widely accepted picture for the Anderson model on the lattice $\ZZ^d$,
for $d=1$ and $d=2$ and any $\lambda\neq 0$, and for  $d \ge 3$ and  large $ \lambda$,
there is only exponential localization, i.e., pure point spectrum with exponentially decaying eigenfunctions.
For $d\geq 3$ and small $\lb\neq 0$, in addition to exponential localization at the spectral edges, the existence of extended states, i.e., absolutely continuous spectrum, 
is expected but not yet proven. 
By now, localization in dimension $d=1$ \cite{GMP,KS1,CKM},  in quasi-one dimensional models (the  strip) \cite{Lac,KLS}, and in  any dimension at the spectral edges  or at high disorder (i.e., large $\lambda$) 
\cite{FS,FMSS,DLS,SW,CKM,DK,K3,AM,A,Wang,Klo} is very well understood.
Localization in dimension $d=2$ at low disorder as well as
absolutely continuous spectrum  in dimensions $d\ge 3$ at small disorder remain open problems.

Localization and delocalization can also be observed by examining the quantum mechanical dynamical behavior, as seen in the spreading of wave packets under the time evolution.
Localization corresponds to effective non-spreading of wave packets  (dynamical localization).
If $d\geq 3$,  diffusive behavior for the spreading of 
wave packets is expected for small $\lambda$.   This is analogous to the     random walk in dimension $d\geq 3$, which is diffusive.

So far, the existence of absolutely continuous spectrum has only been proven for the Anderson model
on the Bethe lattice, the Bethe strip and similar tree like structures.
The first rigorous proof of absolutely continuous spectrum for the Anderson model in the Bethe lattice was obtained by Klein \cite{K1,K2,K9} using a supersymmetric 
transfer matrix method.  These methods were  extended to the Bethe strip in our previous work \cite{KS},  where we proved the  existence of absolutely continuous spectrum in the Bethe strip.

In addition, Klein  showed that the supersymmetric method  also yielded ballistic behavior in the Bethe lattice  \cite{K8}. (Note that the random walk on the Bethe lattice is ballistic.) 
In this paper  we extend these methods to the Bethe strip, proving  ballistic behavior  for the Anderson model in the Bethe strip. 

Different techniques to obtain absolutely continuous spectrum for the Anderson model on the Bethe lattice and similar tree like structures
have been developed in \cite{ASW,FHS,FHH,H,KLW,FHS2,AW}.
The hyperbolic geometry methods of \cite{FHS,H} were extended 
to the Anderson model on a Bethe strip of connectivity $K=2$ and
width $m=2$ in \cite{FHH}. However, up to now the results of \cite{K8} remained the only proof of dynamical delocalization for the Anderson model.

Our proof of absolutely continuous spectrum for the Anderson model on the Bethe strip \cite{KS} used the approach of \cite{K2} combined with the supersymmetric formalism for the strip developed in \cite{KSp2}.   But although the present paper uses the approach of   \cite{K8}, which relied on the methods and  results of \cite{K2}, it does not suffice for us to rely on the methods of \cite{KS}.   The approach is based on a supersymmetric transfer matrix formalism for the Green's function.  In \cite{K8} this lead to the study of certain operators on an  $L^2$-space. On the Bethe strip this is much more complicated, and requires an augmentation of the supersymmetric formalism, with the derivation of new supersymmetric identities and the introduction of new Hilbert and Banach spaces of supersymmetric functions.  This is done in
 Section~\ref{sec-sup-id}; the key results being Theorem~\ref{th-DT} and Corollary~\ref{cor-DT}.
 For the Bethe lattice, i.e., $m=1$,   the Grassmann variables can be integrated out explicitly,
and the Hilbert space $\HH$ (see \eq{defHh}) reduces to a subspace of $L^2(\RR^4)$.  In this case,
the matrix  operator $\T$ (see \eq{eq-def-sT}), a unitary operator on $\HH$ made out of differential operators and the Fourier transform, reduces  to the Fourier transform.  (The differential operators do not appear when $m=1$; this can been seen from the definition \eq{eq-def-sT} and the relation \eqref{eq-DT-FD},  where there are no derivatives on the right hand side when $m=n=1$.)

In this article (as in \cite{KS})   we always make the following assumptions:

\begin{assumptions} $ $
\begin{enumerate} [\upshape (I)]
\item $K \ge 2$,  so $\B$ is not the line $\RR$. 
\item The common probability distribution  $\mu$ of the $\Sym(m)$-valued random variables $\{V(x)\}_{x\in\B}$ has finite  (mixed) moments of all orders. In particular, the characteristic function
of $\mu$,
\begin{equation}
h(M) : = \int_{{\Sym(m)}} {{\e}^{-i\Tr(MV)}d\mu (V)}\quad\text{for}\; M\, \in\,\Sym(m)\;, 
\end{equation}
is a $C^{\infty}$ function on $\Sym(m)$ with bounded derivatives.

\item  Let   $a_{\mathrm{min}} :=a_1\leq a_2\leq \ldots \leq a_m= : a_{\mathrm{max}}$ be the eigenvalues of the  ``free vertical operator'' $A$, and set
\begin{equation} \label{IAK}
I_{A,K} = \bigcap_{i=1}^n (-\sqrt{K}+a_i,\sqrt{K}+a_i) = (-\sqrt{K}+a_{\mathrm{max}},\sqrt{K}+a_{\mathrm{min}}).
\end{equation}
The interval $I_{A,K}$ is not empty, i.e.,
\begin{equation}\label{IAKhyp}
a_{\mathrm{max}}-a_{\mathrm{min}} < 2\sqrt{K}.
\end{equation}  

\end{enumerate}
\end{assumptions}

For a fixed free vertical operator  $A$ one can always obtain \eqref{IAKhyp} by taking $K$ large enough.
To understand the meaning of condition (III), note that $A$ can be diagonalized by a unitary transformation  and
the unperturbed operator $H_0$ can be rewritten as a direct sum of shifted Laplacians on the Bethe lattice (see \cite{KS}).
It follows that  the spectrum of $H_0$ is the union of the spectra of these shifted Laplacians, i.e.,  $\sigma(H_0)=\bigcup_{i=1}^n [-\sqrt{K}+a_i, \sqrt{K}+a_i]$. The interval
$I_{A.K}$ is  simply the interior of the intersection of the spectra of these shifted Laplacians, and condition (III) says that they all overlap.

Let us denote the standard basis elements of $\ell^2(\B,\CC^m)$ by $|x,k\rangle$ for $x\in \B$ and $k\in\{1,\ldots,m\}$,
i.e., $u=|x,k\rangle \in \ell^2(\B,\CC^m)$ is the function $u(y)=\delta_{x,y} e_k$ where
$e_k$ is the $k$-th standard basis vector of $\CC^m$.
A measure for the spread of a wave packet localized at $(x,j) \in \B\times\I$ is given by the square mean displacement
\begin{equation}
 r_{\lb,x,j}^2(t) := \sum_{y\in \B} \sum_{k=1}^m [d((x,j),(y,k))]^2 \left|\langle y,k | e^{-itH_\lb} | x,j \rangle \right|^2,
\end{equation}
where $d((x,j),(y,k))$ denotes the distance between	 the sites$(x,j)$ and $(y,k)$.
For an Anderson model on a product graph $\B\times\G$ this distance would be
$d((x,j),(y,k))=d(x,y)+d(j,k)$ where $d(j,k)$ is the distance between the vertices $j$ and $k$ on the graph $\G$. For a Wegner orbital model
one would choose $d((x,j),(y,k))=d(x,y)$. In any case, we will use $d(x,y)$ as a lower bound. 
Ballistic motion means $r^2_{\lb,x,j}(t) \sim C t^2$, whereas diffusive behavior means
$r^2_{\lb,x,j}(t) \sim Ct$ for large $t$.
One always has ballistic motion as an upper bound,
\begin{equation}
 r_{\lb,x,j}^2(t) \leq Ct^2,
\end{equation}
for some constant $C$ not depending on $x$ and $j$.

In order to show ballistic motion at least for some $|x,j\rangle$ we will consider the sum over $j$ at
some arbitrary site of $\B$  which we will call the origin and denote by $0 $.
Furthermore we set $|x|=d(0,x)$ (which is not a norm) for $x\in \B$ and define
\begin{equation}
 r_\lb^2(t):= \sum_{x\in \B} \sum_{j,k=1}^m |x|^2 \left|\langle x,k | e^{-itH_\lb} | 0,j \rangle \right|^2\;.
\end{equation}
Note that
\begin{equation}\label{eq-sumj}
 \sum_{j=1}^m r_{\lb,0,j}^2(t) \geq r_\lb^2(t)\;.
\end{equation}
\begin{theorem} \label{theo-main}
 For sufficiently small $\lb$ we have
\begin{equation}\label{mainest}
 \liminf_{t\to\infty} \frac{1}{t^3} \int_0^t \E(r^2_\lb(s))\,ds > 0\;.
\end{equation}
In particular, this implies
\begin{equation}\label{mainest2}
 \limsup_{t\to\infty} \E\left(\frac{r^2_\lb(t)}{t^2}\right) > 0 \qtx{and} 
\IP\left(\limsup_{t\to\infty} \frac{r_\lb^2(t)}{t^2} > 0\right) > 0,
\end{equation}
and hence it follows from ergodicity that
\begin{equation}\label{mainest3}
 \IP\left(\limsup_{t\to\infty} \frac{r^2_{\lb,x,j}(t)}{t^2} > 0 \qtx{for some} (x,j)\in\B\times\I \right) = 1\;.
\end{equation}
\end{theorem}

We only need to prove \eq{mainest}, since \eq{mainest2} and \eq{mainest3} are  consequences of \eq{mainest} 
and \eqref{eq-sumj} as shown in \cite{K8}. To prove \eq{mainest}, we start by reformulating the problem  in terms of the matrix valued Green's function.

Given  $x,y \in \B$, $z=E+i\eta$ with $E\in\RR$ and $\eta>0$,  the matrix valued Green's function $G_\lb(x,y;z)$ is the  $m\times m$ matrix 
with entries
\begin{equation}
 \left[G_\lb(x,y;z)\right]_{j,k} :=
\langle x,j | (H_\lb -z)^{-1} | y,k \rangle\;.
\end{equation}
Using the spectral theorem and Plancherel's theorem, as in \cite[Lemma~A.2]{K8}, we obtain
\begin{equation} \label{eq-Tb1}
 \int_0^\infty e^{-\eta t} \E(r_\lb^2(t))\,dt =
\frac{1}{2\pi} \int_{-\infty}^\infty \left\{ \sum_{x\in\B} |x|^2 
\E\left(\Tr \left(\left |G_\lb(0,x;E+i \tfrac{\eta}{2})\right |^2 \right)\right) \right\}\, dE\,.
\end{equation}
Similarly to \cite[eq. (2.3)]{K8},  we also have
\begin{equation} \label{eq-Tb2}
 \int_{-\infty}^\infty \left\{ \sum_{x\in\B} |x|^2 
\E\left(\Tr \left(\left |G_\lb(0,x;E+i \tfrac{\eta}{2})\right |^2 \right)\right) \right\}\, dE \leq
\frac{4\pi m^2}{\eta^3} \left\| \tfrac12 \Delta \right\|^2.
\end{equation}

In view of  \eqref{eq-Tb1} and \eqref{eq-Tb2}, Theorem~\ref{theo-main} is a consequence  of the  following theorem using the  Tauberian Theorem given in \cite[Theorem 10.3]{Simon}  (note that the proof is also valid for
\mbox{lim inf}).

\begin{theorem} \label{theo-main1}  For sufficiently small $\lb$ we have
\begin{equation} \label{eq-lim-eta}
\liminf_{\eta \downarrow 0} \eta^3 \int_{-\infty}^{\infty} \left\{ \sum_{x \in \B} |x|^2 
\E\left( \Tr\left (\left|G_\lb\, (0,x; E +i\eta\right)|^2\right) \right) \right\}\,dE  \; > \; 0 \;.
\end{equation}
More precisely,
there exists $\lambda_0 > 0\,$, such that for any $\lambda$ with $|\lambda| < \lambda_0$
we can find energies
$E_{\lambda}^\pm \in I_{A,K}=(-\sqrt{K}+a_{{\rm max}},\sqrt{K}+a_{{\rm min}})$, with 
\mbox{$\lim_{\lambda \to 0} E_{\lambda}^-= -\sqrt{K}+a_{{\rm max}}$} and 
\mbox{$\lim_{\lambda \to 0} E_{\lambda}^+= +\sqrt{K}+a_{{\rm min}}$},
 such that $H_{\lambda}$ has purely absolutely continuous spectrum  in the interval 
\mbox{$\; I_\lb =( E_{\lambda}^- , E_{\lambda}^+)$,}
and 
\begin{equation}
\liminf_{\eta \downarrow 0}\; \eta^3 \, \sum_{x \in \B} |x|^2 \,\E (|G_{\lb}\, (0,x; E +i\eta)|^2  )
 > 0 \quad\mbox{for all}\quad E \in I_\lb \;.  \label{eq-eta3} 
\end{equation}
\end{theorem}

Equation \eqref{eq-lim-eta} follows immediately from \eqref{eq-eta3} by Fatou's lemma. 
The fact that we find purely absolutely continuous spectrum for small $\lb$ was proved in \cite{KS}, so we only have 
to prove \eqref{eq-eta3}.

The paper is organized as follows. In Section~\ref{sec-super} we introduce the basic supersymmetric formalism, reviewing the definitions and notation we used   in \cite{KS}.
Section~\ref{sec-sup-id} contains the new supersymmetric identities required for this work; see Theorem~\ref{th-DT} and Corollary~\ref{cor-DT}. In Section~\ref{sec-avgreen} 
we use these identities to rewrite the trace
of the averaged matrix valued Green's function, $\E\left(\Tr( |G_\lb(0,x;z)|^2)\right)$, in a convenient form; see Proposition~\ref{prop-GGVv}.  We then finish the proof of Theorem~\ref{theo-main1}   in Section~\ref{sec-proof}.

\section{Supersymmetric methods \label{sec-super}}


The supersymmetric formalism described in this section can be found in more detail in
\cite{B,E,K,KSp2,KLS2,KS}.  We review  the definitions and notation (we mostly  use the same notation as in \cite{KS}) for the reader's convenience.

\subsection{Basic definitions}

By  $\{\psi_{k,\ell}, \overline{\psi}_{k,\ell}; \; k=1,\ldots,m,\,\ell=1,\ldots,n \}$, where  $m,n \in \NN$,  we denote
$2mn$ independent Grassmann variables.  They all anti-commute and are the generators of a
Grassmann algebra isomorphic to $\Lambda^{2mn}(\RR)$,  given by the free algebra over $\RR$ generated by these symbols modulo the ideal generated by the anti-commutators 
$$
\psi_{i,j} \psi_{k,\ell} + \psi_{k,\ell} \psi_{i,j} , \quad  \overline{\psi}_{i,j} \overline{\psi}_{k,\ell} + \overline{\psi}_{k,\ell} \overline{\psi}_{i,j},  \quad \overline{\psi}_{i,j} {\psi}_{k,\ell} + {\psi}_{k,\ell} \overline{\psi}_{i,j} ,
$$
where $ i,k=1,\ldots,m$ and $j,\ell=1,\ldots,n$.
This  finite dimensional algebra  will be denoted by $\Lambda(\BPsi)$, where 
 $\BPsi=(\overline{\psi}_{k,\ell},\psi_{k,\ell})_{k,\ell}$.
The subset of one forms (linear combinations of the generators) is  $\Lambda^1(\BPsi)$.
The complexification of $\Lambda(\BPsi)$ is
$\Lambda_\CC(\BPsi)=\CC\otimes_\RR \Lambda(\BPsi)$.
Sometimes we will also  add and multiply expressions from different Grassmann algebras
$\Lambda(\BPsi)$ and $\Lambda(\BPsi')$; these  are to be understood as sums and products in the Grassmann algebra
 $\Lambda(\BPsi,\BPsi')$,  generated by the entries of $\BPsi$ and $\BPsi'$ as independent
Grassmann variables.

A supervariable is an element of $\RR^2\times\Lambda^1(\BPsi)\times\Lambda^1(\BPsi)$.
We introduce variables $\varphi_{k,\ell} \in \RR^2$ and consider the supervariables  
$\phi_{k,\ell}=(\varphi_{k,\ell},\overline{\psi}_{k,\ell},\psi_{k,\ell})$.
The collection $\BPhi=(\phi_{k,\ell})_{k,\ell}$
will be called a $m \times n$ supermatrix.
More generally, an $m\times n$ matrix 
$\tilde\BPhi=(\tilde\varphi_{k,\ell},\overline{\tilde{\psi}}_{k,\ell},\tilde\psi_{k,\ell})_{k,\ell}\,\in \,
\left[\RR^2\times\Lambda^1(\BPsi)\times\Lambda^1(\BPsi)\right]^{m\times n}$
will be called a supermatrix  if all the 
appearing one forms $\overline{\tilde{\psi}}_{k,\ell},\tilde\psi_{k,\ell}$, $k=1,2\ldots,m$ and $\ell=1,2\ldots,n$, are linearly independent. Supermatrices $(\BPhi_i)_i$ are said to be independent if
$\BPhi_i\in\Ll_{m,n}(\BPsi_i)$ for all $i$, and all the entries of the different
$\BPsi_i$ are independent Grassmann variables.

The collection of all supermatrices is a dense open subset of the vector space
$\left[\RR^2\times\Lambda^1(\BPsi)\times\Lambda^1(\BPsi)\right]^{m\times n}$
 and will be denoted by $\Ll_{m,n}(\BPsi)$, or just $\Ll_{m,n}$.
Linear maps defined on $\Ll_{m,n}(\BPsi)$ have to be understood as restrictions of linear maps 
defined on $\left[\RR^2\times\Lambda^1(\BPsi)\times\Lambda^1(\BPsi)\right]^{m\times n}$.

We also consider  matrices $\bvp=(\varphi_{k,\ell})_{k,\ell}$ with entries in $\RR^{2}$. Writing each entry
$\varphi_{k,\ell}$   as a row vector, $\bvp$ may be considered
as $m\times 2n$ matrix {with real entries}. 
Similarly, one may consider $\BPsi$ as $m\times 2n$ matrix with entries in $\Lambda^1(\BPsi)$.
With all these notations one may write $\BPhi=(\bvp,\BPsi)$,  splitting a
supermatrix into its real and Grassmann-variables parts.

For supervariables $\phi_1=(\varphi_1,\overline{\psi}_1,\psi_1)$ and $\phi_2=(\varphi_2,\overline{\psi}_2,\psi_2)$
we define
\begin{equation}
 \label{eq-dot-supervec}
\phi_1 \cdot \phi_2 : = \varphi_1\cdot\varphi_2+\tfrac12(\overline{\psi}_1 \psi_2 + \overline\psi_2 \psi_1)\;.
\end{equation}
For a supermatrix  $\BPhi$,  $\Phi_k=(\phi_{k,\ell})_{\ell=1\,\ldots,n}$  denotes its $k$-th row vector. Given
two supermatrices $\BPhi$ and $\BPhi'$, we set
\begin{align} \label{eq-dot-rowvec}
\Phi'_j \cdot \Phi_k &  : = \sum_{\ell=1}^n \phi'_{j,\ell} \cdot \phi_{k,\ell}\qtx{for} j,k=1,2,\ldots,m,\\
\label{eq-dot-supermatrix}
 \BPhi' \cdot \BPhi &: = 
\sum_{k=1}^m \Phi_k' \cdot \Phi_k = 
\sum_{k=1}^m \sum_{\ell=1}^n \phi'_{k,\ell}\cdot\phi_{k,\ell}\;.
\end{align}

Given a supermatrix $\BPhi=(\bvp,\BPsi)$, 
where $\bvp\in\RR^{m\times 2n}$ and 
$\BPsi \in \Lambda^1(\BPsi)^{m\times 2n}$, we introduce the $m\times m$ matrix $\BPhi^\ot$ with entries in $\Lambda(\BPsi)$ by
\begin{align}\label{eq-supermatrix-tensor}
(\BPhi^{\od 2})_{j,k} &:=  \Phi_j \cdot \Phi_k = \sum_{\ell=1}^n \left\{\varphi_{j,\ell} \cdot \varphi_{k,\ell}+\tfrac12(\overline \psi_{j,\ell}\psi_{k,\ell}+\overline \psi_{k,\ell}\psi_{j,\ell}) \right\}\\
\nonumber 
&=\sum_{\ell=1}^n \left\{\varphi_{j,\ell} \cdot \varphi_{k,\ell}+ \left[ \begin{matrix} \,\overline{\psi}_{j,\ell} & \psi_{j,\ell} \end{matrix}\right]
\left[\begin{matrix} \,0 & \tfrac12 \\ -\tfrac12 & 0 \end{matrix}\right]
\left[\begin{matrix} \overline{\psi}_{k,\ell} \\ \psi_{k,\ell} \end{matrix}\right]\right\}\;.
\end{align}
It follows that
\begin{equation} \label{eq-BPhi2}
\BPhi^\ot = \bvp^\ot\,+\,\BPsi^\ot\,, \qtx{with}
\bvp^\ot:=\bvp\bvp^\top\;\text{and}\;  \BPsi^\ot:=\BPsi J \BPsi^\top\; ,
\end{equation}
where  $J$ is the $2n\times 2n$ matrix consisting of $n$ blocks 
$\left[\begin{smallmatrix} \,0 & \frac12 \\ -\frac12 & 0 \end{smallmatrix}\right]$ along the diagonal.  Note that  
$\Upsilon^\top$ will always denote the transpose of the matrix $\Upsilon$, whose entries may be elements of  a Grassmann algebra.

Given a complex $m\times m$ matrix $B$, supermatrices $ \BPhi, \,\BPhi'$, and  matrices $\bvp', \bvp\in \RR^{m\times 2n}$, we define
 \begin{align}\label{eq-supermatrix-prod}
\BPhi'\cdot B \BPhi& : = 
\sum_{j,k=1}^m \sum_{\ell=1}^n B_{j,k} \phi'_{j,\ell} \cdot \phi_{k,\ell} =
\sum_{j,k=1}^m B_{j,k} \Phi'_j \cdot \Phi_k\
\in\;\Lambda_\CC(\BPsi) \; ,\\
\bvp'\cdot B \bvp &:= \sum_{j,k=1}^n \sum_{\ell=1}^n B_{j,k} \varphi'_{j,\ell} \cdot \varphi_{k,\ell} = \Tr((\bvp')^\top B \bvp )  \;\in\;\CC \; .
\end{align}
Note that $\BPhi\cdot B \BPhi = \Tr(B \BPhi^\ot)$.

These definitions may be memorized as follows: If $n=1$, $\BPhi$ is a column vector indexed by $k$, 
 $B\BPhi$ is the matrix vector product, and $\BPhi'\cdot B\BPhi$ is the dot product of vectors of supervariables.
For general $n$ the supermatrix  $\BPhi$ has columns indexed by $\ell=1,2,\ldots,n$, ``the $n$ replicas",  and in all definitions 
of dot products
there is an additional sum over this index.

A complex superfunction with respect to $\Lambda(\BPsi)$ is a function 
$F:\RR^{m\times 2n}\to \Lambda_\CC(\BPsi)$.
Let $\beta_{i} \in \Lambda(\BPsi)$ for $i\in\{1,\ldots,2^{2mn}\}$ 
be a basis {for} $\Lambda(\BPsi)$ over $\RR$. 
Each $\beta_{i}$ is a polynomial in the entries of $\BPsi$ (we required the entries of $\BPsi$ to be independent)
and $F$ is of the form  
\begin{equation} \label{eq-superfct-expand}
F(\bvp) = \sum_{i=1}^{2^{2mn}}\, F_i(\bvp)\, \beta_i\;,\quad\text{where}\quad F_i\,:\,\RR^{m\times 2n}\,\to\,\CC\;.
\end{equation} 
We interpret this as a function $F(\BPhi)$ where 
$\BPhi=(\bvp,\BPsi)$. In this sense the map $\BPhi\mapsto  \Tr(B \BPhi^\ot)$, where  $B$ is a complex $m\times m$ matrix,  is  a 
superfunction.
Similarly, we  can define superfunctions $F(\BPhi_1,\ldots,\BPhi_k)$ of $k$  independent supermatrices
using the Grassmann algebra $\Lambda((\BPsi_j)_{j\in\{1,\ldots,k\}})$.
We write $F\in\Ss(\Ll_{m,n})$, or $F\in C^\infty(\Ll_{m,n})$, 
if for all $i$ we have  
$F_i\in\Ss(\RR^{m\times 2n})$, the Schwartz space, or $F_i\in C^\infty(\RR^{m\times 2n})$, respectively.

We define the integral over the Grassmann variables in the following way.
For a fixed pair $k,\ell$ we  write $F=F(\BPhi)$ as  $F=F^{k,\ell}_0+F^{k,\ell}_1\overline{\psi}_{k,\ell}+F^{k,\ell}_2\psi_{k,\ell}+
F^{k,\ell}_3\overline{\psi}_{k,\ell}\psi_{k,\ell}$
where the $F^{k,\ell}_i$ are superfunctions not depending on $\overline{\psi}_{k,\ell}$ and $\psi_{k,\ell}$.
Then
\begin{equation}
\int F\,  d\overline{\psi}_{k,\ell}\,d\psi_{k,\ell}   := 
   -F^{k,\ell}_3 \;.
\end{equation}
If all functions $F_i$ in the expansion \eqref{eq-superfct-expand} are in $L^1(\RR^{m\times2n})$,  we say that
$F\in L^1(\Ll_{m,n})$ and define the supersymmetric integral by
\begin{equation}\label{eq-def-DPhi}
\int F(\BPhi)\;D\BPhi = \frac 1 {\pi^{mn}}
\int\,F(\BPhi)\; \prod_{k=1}^m \prod_{\ell=1}^{n} d^2 \varphi_{k,\ell}\;d\overline{\psi}_{k,\ell}\,d\psi_{k,\ell}\;.
\end{equation}


\subsection{Differential operators and supersymmetric functions}

We now recall the notion of smooth supersymmetric functions  and introduce certain differential operators; we refer to \cite{KS} for details.
We will use the notation of \cite{KS} except for some small sign deviations that will be explicitly pointed out.

We start by introducing  convenient notation for Grassmann monomials. Recall $\I=\{1,\ldots,m\}$;   we will  denote  the set of subsets of $\I$ by
$\PP(\I)$.
Given $(\bar a, a)\in  (\PP(\I))^2=\PP(\I)\times\PP(\I)$
and $\bar a=\{\bar k_1,\ldots, \bar k_c\},\;a=\{k_1,\ldots, k_d\}$, both ordered ({ i.e.}, $\bar k_i < \bar k_j$
and $k_i<k_j$ if $i<j$),  we set 
\begin{equation}\label{Psiaal}
\Psi_{\bar a, a,\ell} :=  \left(\prod_{j=1}^{|\bar a|}\;\overline{\psi}_{\bar k_j, \ell}\,\right)\left(\prod_{j=1}^{|a|}
\psi_{k_j,\ell}\,\right)\;,
\end{equation}
using the conventions $\prod_{j=1}^c \psi_j=\psi_1 \psi_2 \cdots\, \psi_c$ for non-commutative products and
$\prod_{j=1}^0 \psi_j = 1$.
In particular, $\Psi_{\emptyset,\emptyset,\ell}=1$.

An important subset of $(\PP(\I))^2$  is
$\Pp$, the set of pairs $(\bar a, a)$ of subsets of $\I$ with the same cardinality, 
i.e.,
\beq
\Pp=\left\{(\bar a, a)\;: \bar a, a \subset \I\,,\,|\bar a|=|a|\right\}\;.
\eeq
More generally, for each $k\in[-m,m]\cap\ZZ$ we define
\begin{equation}
 \Pp_{k}:=\{(\bar a,a)\;:\bar a, a \subset \I\,,\, |\bar a|=|a|+k\}\;,
\end{equation}
so, in particular, $\Pp=\Pp_{0}$. (A note of caution: we used $\Pp_m$ for $\Pp=\Pp_{0}$
in \cite{KS}.)

For  $(\bar a,a)\in\Pp$ we set 
\begin{equation}\label{eq-P-old}
\Psi^{(\ell)}_{\bar a, a} :=  \prod_{j=1}^{|a|}\;\left(\,\overline{\psi}_{\bar k_j, \ell}\,\psi_{k_j,\ell}\,\right),
\qtx{with the convention}  \Psi^{(\ell)}_{\emptyset,\emptyset} := 1\;.
\end{equation}
Note that these Grassmann monomials are slightly different from the ones we defined in \eq{Psiaal}.
In fact, counting transpositions we get
\begin{equation}\label{eq-psi-relation}
 \Psi_{\bar a, a, \ell} \;=\; (-1)^{\frac{|a|(|a|-1)}{2}} \; \Psi^{(\ell)}_{\bar a, a} \qtx{for} (\bar a,a)\in\Pp\;.
\end{equation}
Given a pair of $n$-tuples of subsets of $\I$, $(\bar\aaa,\aaa) \in (\PP(\I))^n\times(\PP(\I))^n $ with $\bar\aaa=(\bar{a}_1,\ldots,\bar{a}_n)$ and 
$\aaa=(a_1,\ldots,a_n)$, we 
set
\begin{equation} \label{eq-Psi-a-a}
 \Psi_{\bar\aaa,\aaa}:=
\prod_{\ell=1}^n \Psi_{\bar a_\ell,a_\ell, \ell}\;.
\end{equation}

An important subset of $(\PP(\I))^n\times(\PP(\I))^n$ is given by the set
\begin{equation}
\Pp^n:=\{(\bar\aaa,\aaa)\in (\PP(\I))^n\times(\PP(\I))^n\, :\, (\bar a_\ell, a_\ell)\in\Pp\;\text{for}\;\ell=1,\ldots,n\;\} \;.
\end{equation}
This set is canonically isomorphic to the cartesian product $(\Pp)^{\times n}$, justifying the  notation.
However, we will use  $n$ as an upper index to  indicate that we deal with $n$-tuples of sets, and, given  $k\in[-m,m]\cap\ZZ$,
set
\begin{equation}
 \Pp^n_k:=\{(\bar\aaa,\aaa)\in(\PP(\I))^n\times(\PP(\I))^n\, :\, (\bar a_1,a_1)\in\Pp_k,\,
(\bar a_\ell,a_\ell)\in\Pp\;\text{for}\;\ell=2,\ldots,n \}.
\end{equation}
If $(\bar\aaa,\aaa)\in\Pp^n_k$ the sets $\bar a_\ell$ and $ a_\ell$ have the same cardinality except 
for possibly $\ell=1$, where  $|\bar a_1|=| a_1|+k$. Note that $\Pp^n_0=\Pp^n$.
In view of \eqref{eq-psi-relation}, for $\aaa\in(\PP(\I))^n$ we define
\begin{equation}
 \sgn(\aaa)=\prod_{\ell=1}^n (-1)^{\frac{| a_\ell|(| a_\ell|-1)}{2}}\;,
\end{equation}
getting
\begin{equation}\label{eq-rel-defs}
 \prod_{\ell=1}^n \Psi^{(\ell)}_{\bar a_\ell, a_\ell}=\sgn(\aaa)\Psi_{\bar \aaa,\aaa} \qtx{for} (\bar\aaa,\aaa)\in\Pp^n\;.
\end{equation}
(Note: the left hand side of \eqref{eq-rel-defs} corresponds to the definition of $\Psi_{\bar\aaa,\aaa}$ 
for $(\bar\aaa,\aaa)\in\Pp^n$ in \cite[eq.(2.25)]{KS}.)

The counterpart to these Grassmann monomials are differential operators acting on functions defined on  $\Sym^+(m)$, the space of  non-negative, real, symmetric $m\times m$ matrices. 
From now on we assume $n\geq \frac{m}{2}$, so that the map $\bvp \in \RR^{m\times 2n}\mapsto\bvp^\ot=\bvp\bvp^t\in \Sym^+(m)$  is surjective.

Let $C^\infty(\Sym^+(m))$ denote the set of continuous functions $f$ on $\Sym^+(m)$ 
which are $C^\infty$ on the interior of $\Sym^+(m)$. We let
 $\partial_{j,k}$ denote the partial derivative with respect to the $j,k$-entry of the
symmetric matrix, {\it i.e.} $\partial_{j,k} f(M)=\frac{\partial}{\partial M_{k,j}}\; f(M)$ for  $f\in C^{\infty}(\Sym^+(m))$.
 Note that $\partial_{j,k}=\partial_{k,j}$.
We also set  $\tilde \partial_{j,k} = \frac12 \partial_{j,k}$ for $j\neq k$ and  $\tilde \partial_{j,j}=\partial_{j,j}$.

Given $(\bar a, a)\in\Pp$ with $a\not=\emptyset$, $\bar a=\{\bar k_1,\ldots,\bar k_c\}$ and $ a=\{k_1,\ldots, k_c\}$, both ordered,
we define the matrix-differential operator
\begin{equation}
 \bpart_{\bar a, a}:=\begin{pmatrix}
       \tilde \partial_{\bar k_1,k_1} & \cdots & \tilde \partial_{\bar k_1,k_c}  \\
        \vdots & \ddots & \vdots \\
	\tilde \partial_{\bar k_c,k_1} & \cdots & \tilde \partial_{\bar k_c, k_c} \end{pmatrix}\;.  
\end{equation}
Furthermore we set $D_{\emptyset,\emptyset}$ to be the identity operator and
\begin{equation}
 D_{\bar a,a} := \det (\bpart_{\bar a,a})  \qtx{if}  a\not= \emptyset\,.
\end{equation}
Note that $\bpart_{\bar a,a}=\bpart_{a,\bar a}^\top$, and hence $D_{\bar a,a}=D_{a,\bar a}$.
In the special case when $\bar a=a=\I$, we set
\begin{equation}
\bpart:=\bpart_{\I,\I}\qtx{and}
\delta:=\det(\bpart) = D_{\I,\I}\;.
\end{equation}
Furthermore, for $(\bar a, a)\in \Pp$ we set $(\bar a^\rc, a^\rc):=(\I\setminus \bar a\,, \I\setminus a)\in\Pp$.
Note that the $(j,k)$-entry of $\bpart$ is $\bpart_{\{j\},\{k\}}=D_{\{j\},\{k\}}$. The cofactor is given by
$(-1)^{j+k} D_{\{j\}^\rc,\{k\}^\rc}$. 
The transpose of the cofactor matrix of $\bpart$ will be denoted by $\bslp$, i.e.,
\begin{align}
\bslp&:=\left((-1)^{j+k}D_{\{k\}^\rc,\{j\}^\rc} \right)_{j,k \in \I} \\
&=
\begin{pmatrix}
D_{\{1\}^\rc,\{1\}^\rc} & - D_{\{2\}^\rc,\{1\}^\rc} & \ldots & (-1)^{m+1} D_{\{m\}^\rc,\{1\}^\rc} \\
-D_{\{1\}^\rc,\{2\}^\rc} & D_{\{2\}^\rc,\{2\}^\rc} & \ldots &  (-1)^{m} D_{\{m\}^\rc,\{2\}^\rc}\\       
\vdots & \vdots & \ddots & \vdots \\ 
(-1)^{m+1} D_{\{1\}^\rc,\{m\}^\rc} & (-1)^m D_{\{2\}^\rc,\{m\}^\rc} & \ldots &  D_{\{m\}^\rc,\{m\}^\rc}\\       
\end{pmatrix} \notag\;.
\end{align}
With these definitions we have
\begin{equation} \label{eq-DD=d}
\bpart \,\bslp=\bslp\,\bpart= \det (\bpart) \one=\dlt:=
\begin{pmatrix}
 \delta & & \!\!\!\!\!\!\!\!\! \mathbf{0} \\
& \ddots & \\
 & \!\!\!\!\!\!\!\!\!\!\!\! \mathbf{0} & \delta
\end{pmatrix},
\end{equation}
where $\one=\one_m$ denotes the unit $m\times m$ matrix and $\dlt$ is a diagonal $m\times m$ matrix
with $\delta$ on all diagonal entries.
Furthermore, for $(\bar\aaa, \aaa) \in \Pp^n$ we set
\begin{equation} \label{eq-def-Daa}
D_{\bar \aaa, \aaa} :=  \prod_{\ell=1}^n D_{\bar a_\ell,  a_\ell}\; .
\end{equation}
We have $D_{\bar\aaa,\aaa}=D_{\aaa,\bar\aaa}$.
For $f\in C^\infty(\Sym^+(m))$ and $\det(\bvp^\ot) \neq 0$, a formal Taylor expansion yields
(cf. \cite[eq.(2.26)]{KS})
\begin{equation}\label{eq-super-Taylor}
 f(\BPhi^{\od 2}) = 
\sum_{ (\bar\aaa, \aaa)\,\in\,\Pp^n}
D_{\bar\aaa,\aaa}\; f(\bvp^\ot)\;\sgn(\aaa) \Psi_{\bar\aaa,\aaa}\;.
\end{equation}
Let $C^\infty_n(\Sym^+(m))$ denote
the set of all functions $f\in C^\infty(\Sym^+(m))$ where $\bvp\mapsto D_{\bar\aaa,\aaa} f(\bvp^\ot)$, defined
on the dense open set where $\det (\bvp^\ot)\neq 0$, extends (uniquely) to a
$C^\infty$ function on $\RR^{m \times 2n}$ for all $(\bar\aaa,\aaa)\in \Pp^n$.

\begin{defini}
Let $n\geq\frac{m}{2}$.
\begin{enumerate}[\upshape (i)]
\item The set $SC^\infty(\Ll_{m,n})$ of smooth supersymmetric functions is defined as the set of all smooth
superfunctions $F(\BPhi)$ such that $F(\BPhi)=f(\BPhi^\ot)$ for some
$f\in C^\infty_n(\Sym^+(m))$.
\item
 The set $S\Ss(\Ll_{m,n}):=\Ss(\Ll_{m,n})\cap SC^\infty(\Ll_{m,n})$ denotes the 
supersymmetric Schwartz functions.
\end{enumerate}
\end{defini}

This definition is justified by  \cite[Proposition~2.3]{KS}, which is based on  \cite[Corollary~2.9]{KSp2}.  $SC^\infty(\Ll_{m,n})$ can be identified with $C^\infty_n(\Sym^+(m))$.
Furthermore, if we define the subset $\Ss_n(\Sym^+(m))$ of all functions $f\in C^\infty_n(\Sym^+(m))$
where $\bvp\mapsto D_{\bar\aaa,\aaa} f(\bvp^\ot)$ is a Schwartz function, then
$S\Ss(\Ll_{m,n})$ can be identified with $\Ss_n(\Sym^+(m))$.

Finally let us define some algebraic operations on $(\PP(\I))^n$ which will give some Leibniz type formulas and will be useful later.
Let $\aaa, \bbb \in (\PP(\I))^n$. 
If $ a_\ell\cap b_\ell=\emptyset $ for each $\ell=1,\ldots,n$, then we say $\aaa$ and $\bbb$ are {\em addable} and define
$\ccc=\aaa+\bbb \in (\PP(\I))^n$ by
$c_\ell= a_\ell\cup b_\ell$.
Similarly, if $b_\ell \subset  a_\ell$ for all $\ell$, then
we define $\ccc=\aaa-\bbb$ by $c_\ell =  a_\ell\setminus b_\ell$.
If $\aaa$ and $\bbb$ are addable, then $(\aaa+\bbb)-\bbb=\aaa$. 
Furthermore we denote by $\III$ the $n$-tuple 
where each entry is the full set $\I$, $\III=(\I,\I,\ldots,\I)$. 
Moreover, we define $\aaa^\brc=\III-\aaa$.
We say that $(\bar\aaa,\aaa)$ and $(\bar\bbb,\bbb)\in \PP(\I)^n \times \PP(\I)^n$ are addable if
$\bar\aaa+\bar\bbb$ and $\aaa+\bbb$ are defined by the notion above.
In this case we define $\sgn(\bar\aaa,\aaa,\bar\bbb,\bbb)\in\{-1,1\}$ by
\begin{equation} \label{eq-def-sgn4}
\Psi_{\bar\aaa,\aaa}\,\Psi_{\bar\bbb,\bbb} = 
\sgn(\bar\aaa,\aaa,\bar\bbb,\bbb)\; \Psi_{\bar\aaa+\bar\bbb,\aaa+\bbb}\;.
\end{equation}
(Note of caution: this definition of $\sgn(\bar\aaa,\aaa,\bar\bbb,\bbb)$ differs from \cite[eq.(2.27)]{KS} since our  definition  of
$\Psi_{\bar\aaa,\aaa}$ in \eq{eq-Psi-a-a} is different from \cite[eq.(2.25)]{KS}.)
If $(\bar\aaa,\aaa) \in \Pp^n_j$ and $(\bar\bbb,\bbb)\in\Pp^n_k$ are addable, then
$(\bar\aaa+\aaa,\bar\bbb+\bbb)\in\Pp^n_{j+k}$ and $(\bar\aaa^\brc,\aaa^\brc),\,(\aaa,\bar\aaa)\,\in \Pp^n_{-j}$.

Since the product of two supersymmetric functions is supersymmetric,  for all
$f, g\in {C^\infty_n(\Sym^+(m))}$ and all $(\bar\aaa,\aaa)\in\Pp^n$ we have
\begin{equation} \label{eq-Leibn}
D_{\bar\aaa,\aaa}\,(fg) = 
\sum_{\substack{ (\bar\bbb,\bbb),(\bar\bbb',\bbb')\in\Pp^n\\ \bar\bbb+\bar\bbb'=\bar\aaa\;,
\bbb+\bbb'=\aaa}}\;\frac{\sgn(\bbb)\sgn(\bbb')\sgn(\aaa)}{\sgn(\bar\bbb,\bbb,\bar\bbb',\bbb')}\,
D_{\bar\bbb,\bbb}\, g\,D_{\bar\bbb',\bbb'}\, f\;.
\end{equation}

\subsection{The supersymmetric Fourier transform}

We recall the definition of the supersymmetric Fourier transform.
Given $f\in S\Ss_n(\Sym^+(m))$ we define $Tf\in \Ss_n(\Sym^+(m))$ by
\begin{equation}\label{eq-def-T}
 (Tf)((\BPhi')^{\od 2}) = 
\int e^{\imath \BPhi'\cdot\BPhi}\, f({\BPhi}^{\od 2})\; D\BPhi\; ,
\end{equation}
where we use the fact that the right hand side defines a  supersymmetric function 
\cite{KSp2}.  It follows that  \cite[eq. (2.37)]{KS}
\begin{equation}\label{eq-T-parts}
D_{\bar\aaa,\aaa} (Tf) = 
\tfrac{2^{mn}}{4^{|\aaa|}}\;\sgn(\aaa,\bar\aaa)\;\Ff(D_{\aaa^\brc,\bar\aaa^\brc}\,f) \qtx{for all} (\bar\aaa,\aaa) \in \Pp^n\; .
\end{equation}
Here  $\Ff$ denotes the Fourier transform on $\RR^{m\times 2n}$; we abuse the notation by letting $\Ff f$ denote  
the function in $\Ss_n(\Sym^+(m))$ such that $(\Ff f)(\bvp^\ot)$ is the Fourier transform of the function  $F(\bvp)=f(\bvp^\ot)$.   In addition, 
\begin{equation}\label{eq-def-sgn2}
\sgn(\bar\aaa,\aaa) :=
\int \sgn(\aaa) \Psi_{\bar\aaa,\aaa}\;\sgn(\aaa^\brc)\Psi_{\bar\aaa^\brc,\aaa^\brc}\,D\BPsi
=(-1)^{mn}\frac{\sgn(\aaa)\sgn(\aaa^\brc)\sgn(\III)}{\sgn(\bar\aaa,\aaa,\bar\aaa^\brc,\aaa^\brc)}
\end{equation}
for $(\bar\aaa,\aaa)\in\Pp^n$, where
$D\BPsi = \prod_{k,\ell} d\overline{\psi}_{k,\ell}\,d\psi_{k,\ell}$.
By \eqref{eq-rel-defs} this is the  same definition as \cite[eq.(2.32)]{KS};
the second equality follows from \eqref{eq-def-sgn4} and
\begin{equation}\label{eq-intPI}
\int \Psi_{\III,\III}D\BPsi=(-1)^{mn}\sgn(\III)\;.
\end{equation}
Moreover  we have, as in \cite[eq. (2.39)]{KS},
\begin{equation} \label{eq-T2=1}
T^2f =  TTf = f\; \qtx{for all} f\in{ \Ss_n(\Sym^+(m))}.
\end{equation}

In the special cases when $(\bar\aaa,\aaa)=(\III,\III)$ and $(\bar\aaa,\aaa)=(\III^\brc,\III^\brc) $,  \eqref{eq-T-parts}  yields
\begin{equation} \label{eq-dT-F}
 \delta^n T f = (-2)^{-mn} \Ff f \qtx{and}
T f = (-2)^{mn}  \Ff \delta^n f\;.
\end{equation}
In particular, we have
\begin{equation}
 \Ff = 4^{mn} \delta^n \Ff \delta^n .
\end{equation}

An important special case of \eqref{eq-T-parts} is the following.
For $k\in\I$ let
\begin{equation}
 \lt k \rt:=(\{k\},\emptyset,\ldots,\emptyset)\,\in\,\PP(\I)^n\;,
\end{equation}
then $\Psi_{\lt j \rt,\lt k \rt}=\bar \psi_{j,1} \psi_{k,1}$, $D_{\lt j \rt,\lt k \rt}=D_{\{j\},\{k\}}$ and
$D_{\lt j \rt^\brc,\lt k \rt^\brc}=D_{\{k\}^\rc,\{j\}^\rc} \delta^{n-1}$.
Using the notation of \eqref{eq-P-old} we get\begin{equation}
\bar \psi_{j,1} \psi_{k,1} \Psi^{(1)}_{\{j\}^\rc,\{k\}^\rc}=(-1)^{j+k} \Psi^{(1)}_{\I,\I}
\end{equation}
since one needs $j-1$ transpositions to bring the $\bar\psi$'s to order and $k-1$ transpositions to bring the $\psi$'s to order.
In view of \eqref{eq-rel-defs} this implies
\begin{equation}
 \sgn(\lt j \rt)\Psi_{\lt j \rt,\lt k \rt}\, \sgn(\lt j \rt^\brc) \Psi_{\lt j \rt^\brc,\lt k \rt^\brc}=(-1)^{j+k} \sgn(\III) \Psi_{\III,\III},
\end{equation}
which by \eqref{eq-def-sgn2} and \eqref{eq-intPI} yields
\begin{align}
\sgn(\lt j \rt,\lt k \rt)&=
\int (-1)^{j+k} \,\sgn(\III)\Psi_{\III,\III}D\BPsi = (-1)^{mn} (-1)^{j+k}\;.
\end{align}
Thus, using $\bar\aaa=\lt j \rt$ and $\aaa=\lt k \rt$ in \eqref{eq-T-parts} gives
\begin{equation}
 D_{\{j\},\{k\}} (Tf)= 2^{mn-2} (-1)^{mn} (-1)^{j+k} \Ff( \delta^{n-1} D_{\{k\}^\rc,\{j\}^\rc} f)\;.
\end{equation}
In particular, given  $\f=(f_1,\ldots,f_m)^\top \in [\Ss_n(\Sym^+(m))]^{m\times 1}$,  a column vector of elements of $\Ss_n(\Sym^+(m))$, setting
\beq
T\f= (T f_1,\ldots,T f_m)^\top \qtx{and}
\Ff \f = (\Ff f_1, \ldots, \Ff f_m)^\top,
\eeq
we have 
\begin{equation} \label{eq-DT-FD}
\bpart (T\f)=(-2)^{mn-2}\, \Ff(\dlt^{n-1} \bslp \f)\;.
\end{equation}

As in \cite{KS}, following Campanino and Klein \cite{CK,K,KSp2}, we introduce   norms $\hn  \cdot  \hn_p$ on 
 $ C^\infty_n(\Sym^+(m))$,  with  $p\in [1,\infty]$,  by
\begin{equation} \label{eq-def-norms}
\hn f \hn_p^2\;: =\;
\sum_{(\bar\aaa,\aaa)\in\Pp^n}\; \left\|\,2^{|\aaa|}\;
D_{\bar\aaa,\aaa} \;f\,(\bvp^\ot)\right\|^2_{L^p(\RR^{m\times 2n}, d^{2mn}\bvp)}.
\end{equation}
We  define the Hilbert space $\Hh$ as completion of ${ \Ss_n(\Sym^+(m))}$ with respect to the norm $\hn\cdot\hn_2$.  
The Banach spaces  $\Hh_p$,  $p \in [1,\infty]$, are defined by
\begin{equation}\label{eq-def-spaces}
\Hh_p\ : =\;\{f\;\in\;\Hh\;:\; \|f\|_{\Hh_p}\; :=\;\hn f \hn_2 +\hn f \hn_p\;<\; \infty \}.
\end{equation}
We also define the Banach space $\widetilde{\Hh}_\infty$ as the completion of
\beq
\widetilde{\Hh}_\infty^{(0)}:= \set{f \in C^\infty_n(\Sym^+(m)) \; :\;  \hn f \hn_\infty < \infty}
\eeq
with respect to the norm $\hn\cdot\hn_\infty$.
In view of \eqref{eq-T-parts} and \eqref{eq-def-norms}, as mentioned in \cite{KS}, 
the supersymmetric Fourier transform $T$ extends to $\Hh$ 
as a unitary operator. 

We also want to consider tensor products.
For $g(\bvp^\ot_+,\bvp^\ot_-) \in C^\infty_n(\Sym^+(m))\otimes  C^\infty_n(\Sym^+(m))$ we define the tensor norms
 \begin{equation}
\hnn g \hnn_p^2  \; = 
\sum_{\substack{(\bar\aaa,\aaa)\in \Pp^n\\(\bar\bbb,\bbb)\in \Pp^n}}
\left\| 2^{|\aaa|+|\bbb|}\,D^{(+)}_{\bar \aaa,\aaa}\, D^{(-)}_{\bar\bbb,\bbb} \, g(\bvp_+^\ot, \bvp_-^\ot) \,
\right\|^2_{{L}^p(\bvp_+,\bvp_-)} ,
\end{equation}
where $\|\cdot\|_{L^p(\bvp_+,\bvp_-)}$ denotes the $p$-norm of the $L^p$ space on ${\left(\RR^{m\times 2n}\right)}^2$
in the variables $\bvp_+,\bvp_-$ {with respect to} the Lebesgue measure $d^{2mn} \bvp_+\, d^{2mn} \bvp_-$.
$D^{(+)}_{\bar \aaa,\aaa}$, $D^{(-)}_{\bar\aaa,\aaa}$  denote the differential operator $D_{\bar \aaa,\aaa}$ with respect to  $\bvp^\ot_+$ and
$\bvp^\ot_-$ respectively.
The Hilbert space tensor product $\Kk:=\Hh\otimes\Hh$ is the completion of $\Ss_n(\Sym^+(m))\otimes\Ss_n(\Sym^+(m))$ with respect to 
the norm $\hnn\cdot\hnn_2$. The unitary operator $T$ induces the unitary transformation
\begin{equation}\label{eq-def-Tt}
 \wt:=T\otimes T  \qtx{on} \Kk .
\end{equation}
 As in \cite{KS} we also define the Banach spaces
\begin{equation}
{\Kk}_p  \; = \; \{g \in {\Kk}\;,\;\;\|g\|_{{\Kk}_p}  =  \hnn g\hnn_2 +\hnn g\hnn_p  < \infty \} \qtx{for}  1 \le p \le \infty \;.
\end{equation}
  As before, we also define the the Banach space $\widetilde{\Kk}_\infty$ as the completion of
\beq
\widetilde{\Kk}_\infty^{(0)}:= \set{g \in C^\infty_n(\Sym^+(m))\otimes  C^\infty_n(\Sym^+(m)) \; :\;  \hn g \hn_\infty < \infty}
\eeq
with respect to the norm $\hn\cdot\hn_\infty$.

\begin{remark}  The spaces $\Hh$, $\Hh_p$, $\widetilde{\Hh}_\infty$, $\Kk$, $\Kk_p$, $\widetilde{\Kk}_\infty$,
the supersymmetric Fourier transform $T$, etc., 
all depend on our choice of $n\ge \frac m 2$ for a given $m$.  This dependence on $n$ (and $m$) will be generally omitted.
\end{remark}


\section{More supersymmetric identities \label{sec-sup-id}}
 In this section we derive new supersymmetric identities that are crucial for the extension of the results of   \cite{K8} to the Bethe strip, going beyond the  supersymmetric formalism used in  \cite{KS}

Using the first replica, we define the following Grassmann column vectors
\begin{equation} \label{eq-def-vpsi}
 \vbpsi:=(\overline{\psi}_{1,1}, \overline{\psi}_{2,1}, \ldots, \overline{\psi}_{m,1})^\top\;,\quad
 \vpsi:=(\psi_{1,1},\psi_{2,1},\ldots,\psi_{m,1})^\top\;
\end{equation}
which correspond to the first and second column vector of $\BPsi$.
Even though we only use the first replica (the second index is always one) for these vectors we do not add an index 1 to $\vbpsi$ or $\vpsi$. 
We want to avoid having too many indices later, when we use a corresponding notation for an indexed family of supermatrices. 
Given  $\f=(f_1,\ldots,f_m)^\top \in [\Lambda(\BPsi)]^{m\times 1}$,  a column vector of elements of the Grassmann algebra, we set
\begin{equation}
 \vbpsi \cdot \f := {\vbpsi}^\top \f = \sum_{k=1}^m \overline{\psi}_{k,1} f_k\qtx{and}
\vpsi \cdot \f := {\vpsi}^\top \f = \sum_{k=1}^m {\psi}_{k,1} f_k\; .
\end{equation}
For an $m\times m$ matrix $\bF=(F_{j,k})_{j,k\in\I} \in [\Lambda(\BPsi)]^{m\times m}$ 
the expressions $\bF \vbpsi$ and $\bF \vpsi$ will be understood as  matrix multiplication.
\begin{remark}
Vectors of elements of the Grassmann algebra $\Lambda(\BPsi)$ and vectors of functions will always be considered as
column vectors in matrix products. In particular, the sets $\Lambda(\BPsi)^m$ and $[\Ss_n(\Sym^+(m))]^m$ will be identified with
$[\Lambda(\BPsi)]^{m\times 1}$ and $[\Ss_n(\Sym^+(m))]^{m\times 1}$, respectively.
\end{remark}

Given $\aaa=(a_1,\ldots,a_n) \in (\PP(\I))^n$ and $k\leq|a_1|$ we let $a_{1k}$ denote the
$k$-th smallest element of $a_1$. Similarly, $\bar a_{1k}$ will denote the $k$-th smallest element of $\bar a_1$ for
$\bar\aaa=(\bar a_1,\ldots,\bar a_n)\in(\PP(\I))^n$.
If  $(\bar\aaa,\aaa)\in\Pp_1^n$ and $k=1,2,\ldots, |\bar{a}_1|$, we have $(\bar\aaa-\lt\bar a_{1k}\rt,\aaa)\in\Pp^n$, and for
$(\bar\aaa,\aaa)\in\Pp^n_{-1}$  and $k=1,2,\ldots, |{a}_1|$ we have $(\bar\aaa,\aaa-\lt a_{1k}\rt)\in\Pp^n$.
Given $\f=(f_1,\ldots,f_m)^\top \in [\Ss_n(\Sym^+(m))]^m$, we have
\begin{align} \notag
 {\vbpsi} \cdot \f(\BPhi^\ot) &=
\sum_{k=1}^m \overline{\psi}_{k,1} f_k(\BPhi^\ot)=
\sum_{k=1}^m\sum_{(\bar\bbb,\bbb)\in\Pp^n} D_{\bar\bbb,\bbb} f_k(\bvp^\ot) \sgn(\bbb) \overline{\psi}_{k,1}\Psi_{\bar\bbb,\bbb}\\
&= \sum_{(\bar\aaa,\aaa)\in\Pp_1^n} \sum_{k=1}^{|\bar a_1|} D_{\bar\aaa-\lt\bar a_{1k}\rt, \aaa}\, f_{\bar a_{1k}}(\bvp^\ot) \;\sgn(\aaa) 
(-1)^{k-1} \Psi_{\bar\aaa,\aaa} \label{eq-vexp-1}\;.
\end{align}
The change in the sum is done by the relations $\aaa=\bbb,\; \bar\aaa-\lt\bar a_{1k}\rt=\bar\bbb$. Note that for 
$k\in\bar b_1$ we have $\overline{\psi}_{k,1}\Psi_{\bar\bbb,\bbb}=0$.
Similarly, we obtain
\begin{equation} \label{eq-vexp-2}
 {\vpsi} \cdot \f(\BPhi^\ot)  =
\sum_{(\bar\aaa,\aaa)\in\Pp_{-1}^n} \sum_{k=1}^{| a_1|} D_{\bar\aaa, \aaa-\lt a_{1k}\rt}\, f_{ a_{1k}}(\bvp^\ot) \;\sgn(\bar\aaa) 
(-1)^{k-1+|\bar a_1|} \Psi_{\bar\aaa,\aaa}
\end{equation}

In order to obtain the super Fourier transform of these expression we will expand $e^{i\BPhi\cdot\BPhi'}$ in the Grassmann variables.
\begin{lemma} We have 
\begin{equation}
 e^{i\BPhi\cdot\BPhi'} =
 e^{i\bvp\cdot\bvp'}\!\!\!\!\!\!\sum_{\bar \aaa,\aaa\in\PP(\I)^n}\!\! \left\{ (\tfrac{i}{2})^{|(\bar\aaa,\aaa)|}
\tfrac{\sgn(\aaa)}{\sgn(\bar\aaa)}(-1)^{|\aaa|} s(\bar\aaa,\aaa) \Psi'_{\bar\aaa,\aaa} \Psi_{\aaa,\bar\aaa}
\right\}, \label{eq-exp-PP}
\end{equation}
where
\begin{equation}
|\aaa|:=\sum_{\ell=1}^n | a_\ell| \qtx{for}\aaa\in\PP(\I)^n\;,\quad
|(\bar\aaa,\aaa)|:=|\bar\aaa|+|\aaa|\qtx{for} \bar\aaa,\aaa\in\PP(\I)^n ,
\end{equation}
and
\begin{equation}
s(\bar\aaa,\aaa):=\prod_{1\leq\ell < \ell'\leq n} (-1)^{(|\bar a_\ell|+| a_\ell|)\cdot(|\bar a_{\ell'}|+| a_{\ell'}|)}\;.
\end{equation}
\end{lemma}

\begin{proof} Recalling that for  non-commutative products we always use the convention that the indices are ordered, increasing from left to right, we have
\begin{align}
 &e^{i\BPhi\cdot\BPhi'} =
e^{i\bvp\cdot\bvp'} 
\prod_{k=1}^m \prod_{\ell=1}^n \left[\left(1+\tfrac{i}{2} \bar\psi_{k,\ell} \psi'_{k,\ell} \right)
\left(1+\tfrac{i}{2} \bar\psi'_{k,\ell} \psi_{k,\ell} \right)\right] \\
&\;= 
e^{i\bvp\cdot\bvp'}\!\!\!\!\!\!
\sum_{\bar \aaa,\aaa\in\PP(\I)^n} \!\!\left\{ (\tfrac{i}{2})^{|(\bar\aaa,\aaa)|}
\prod_{\ell=1}^n  \Bigg[\prod_{k\in \bar a_\ell} \bar\psi'_{k,\ell} \psi_{k,\ell}
\prod_{j\in a_\ell} \bar\psi_{j,\ell} \psi'_{j,\ell}\Bigg]
\right\} \notag \\
&\;= 
e^{i\bvp\cdot\bvp'}\!\!\!\!\!\!
\sum_{\bar \aaa,\aaa\in\PP(\I)^n}\!\! \left\{ (\tfrac{i}{2})^{|(\bar\aaa,\aaa)|}
\tfrac{\sgn(\aaa)}{\sgn(\bar\aaa)} 
\prod_{\ell=1}^n  \Bigg[\prod_{k\in \bar a_\ell} \bar\psi'_{k,\ell} \prod_{k\in\bar a_\ell}\psi_{k,\ell}
\prod_{j\in a_\ell} \bar\psi_{j,\ell}\prod_{j\in a_\ell}\psi'_{j,\ell}\Bigg]
\right\} \notag \\ \notag
&\;= 
e^{i\bvp\cdot\bvp'}\!\!\!\!\!\! 
 \sum_{\bar \aaa,\aaa\in\PP(\I)^n}\!\! \Bigg\{ (\tfrac{i}{2})^{|(\bar\aaa,\aaa)|}
\tfrac{\sgn(\aaa)}{\sgn(\bar\aaa)}(-1)^{|\aaa|}\\ & \qquad \qquad  \qquad \qquad \qquad \qquad  \times
 \prod_{\ell=1}^n  \Bigg[\prod_{k\in \bar a_\ell} \bar\psi'_{k,\ell} \prod_{j\in a_\ell}\psi'_{j,\ell}
\prod_{j\in a_\ell} \bar\psi_{j,\ell}\prod_{k\in\bar a_\ell}\psi_{k,\ell}\Bigg]
\Bigg\} \notag \\
&\;= 
e^{i\bvp\cdot\bvp'}\!\!\!\!\!\!
\sum_{\bar \aaa,\aaa\in\PP(\I)^n}\!\! \left\{ (\tfrac{i}{2})^{|(\bar\aaa,\aaa)|}
\tfrac{\sgn(\aaa)}{\sgn(\bar\aaa)}(-1)^{|\aaa|} s(\bar\aaa,\aaa) \Psi'_{\bar\aaa,\aaa} \Psi_{\aaa,\bar\aaa}
\right\} .\notag  \qedhere 
\end{align}
\end{proof}
Note that for $(\bar\aaa,\aaa)\in\Pp^n_k$ and $\ell\geq 2$ we have $|\bar a_\ell|=| a_\ell|$, therefore
\begin{equation}
 s(\bar\aaa,\aaa)=1\qtx{for all} (\bar\aaa,\aaa)\in\Pp^n_k\qtx{and all} k\in[-m,m]\cap\ZZ\;. 
\end{equation}

Now if we combine \eqref{eq-vexp-1} and \eqref{eq-exp-PP},
using $[\frac{i}{2}]^{|(\bar\aaa,\aaa)|}=2i \frac{(-1)^{|\aaa|}}{4^{|\bar\aaa|}}$ 
for $(\bar\aaa,\aaa)\in\Pp_1^n$ as well as \eqref{eq-def-sgn4}, \eqref{eq-intPI} and \eqref{eq-T-parts} for the integral over $D\BPsi$,
then we obtain
\begin{align}  \label{eq-vec-int}
 & \int e^{i\BPhi\cdot\BPhi'} \vbpsi\cdot \f(\BPhi^\ot)
\;D\BPhi\\
& \qquad  = \sum_{(\bar\aaa,\aaa)\in\Pp^n_1} \Bigg\{
\frac{\sgn(\aaa)\sgn(\bar\aaa^\brc)\sgn(\III)}{\sgn(\bar\aaa)\sgn(\aaa,\bar\aaa,\aaa^\brc,\bar\aaa^\brc )}(-1)^{mn} \Psi'_{\bar\aaa,\aaa} 
\notag \\
& \qquad\qquad\qquad\qquad  \qquad \qquad   \times 2i \frac{2^{mn}}{4^{|\bar\aaa|}}
\sum_{k=1}^{|a^\rc_1|} (-1)^{k-1} \big(\Ff D_{\aaa^\brc-\lt a^\rc_{1k}\rt, \bar\aaa^\brc}
f_{a^\rc_{1k}}\big) (\bvp'^\ot)
\Bigg\} \notag \\ 
&\qquad  = 2i  \sum_{(\bar\aaa,\aaa)\in\Pp^n_1} \bigg\{  \sum_{k=1}^{|a^\rc_1|} (-1)^{a^\rc_{1k}-k+1}
 D_{\bar\aaa, \aaa+\lt a^\rc_{1k}\rt} \,T f_{a^\rc_{1k}} (\bvp'^\ot) \bigg\}  \sgn(\aaa) \Psi'_{\bar\aaa,\aaa}\;. \notag
\end{align}
Note that $a^\rc_{1k}=(a^\rc_1)_k$ denotes the $k$-th smallest element of the set $a^\rc_1=\I\setminus a_1$.
To get the sign in the last equation we used
\begin{align}
& \frac{\sgn(\bar\aaa)\sgn(\bar\aaa^\brc)\sgn(\III)}{\sgn(\aaa,\bar\aaa,\aaa^\brc,\bar\aaa^\brc )}(-1)^{mn}
\sgn(\aaa+\lt a^\rc_{1k}\rt,\bar\aaa)(-1)^{k-1}  \\
& \qquad\qquad  \qquad \qquad \qquad =\frac{\sgn(\aaa+\lt a^\rc_{1k}\rt,\bar\aaa,\aaa^\brc-\lt a^\rc_{1k}\rt,\bar\aaa^\brc)}{\sgn(\aaa,\bar\aaa,\aaa^\brc,\bar\aaa^\brc )}
(-1)^{k-1}, \notag
\end{align}
and the fact that for $(\aaa,\bar\aaa)\in\Pp_{-1}^n$ we have
\begin{align}
 \Psi_{\aaa,\bar\aaa}\Psi_{\aaa^\brc,\bar\aaa^\brc}&=(-1)^{k-1} \Psi_{\aaa,\bar\aaa} \overline{\psi}_{a^\rc_{1k},1}
\Psi_{\aaa^\brc-\lt a^\rc_{1k}\rt,\bar\aaa^\brc}=(-1)^k \overline{\psi}_{a^\rc_{1k},1}
\Psi_{\aaa,\bar\aaa} \Psi_{\aaa^\brc-\lt a^\rc_{1k}\rt,\bar\aaa^\brc} \notag \\
&= (-1)^k (-1)^{a^\rc_{1k}-k} \Psi_{\aaa+\lt a^\rc_{1k}\rt,\bar\aaa} \Psi_{\aaa^\brc-\lt a^\rc_{1k}\rt,\bar\aaa^\brc},
\end{align}
implying 
\begin{equation}
 \frac{\sgn(\aaa+\lt a^\rc_{1k}\rt,\bar\aaa,\aaa^\brc-\lt a^\rc_{1k}\rt,\bar\aaa^\brc)}
{\sgn(\aaa,\bar\aaa,\aaa^\brc,\bar\aaa^\brc )}=(-1)^{a^\rc_{1k}}\;.
\end{equation}

Since for $(\bar\aaa,\aaa)\in\Pp^n_{-1}$ we have
 \beq
 \left(\frac{i}{2}\right)^{|(\bar\aaa,\aaa)|}=-2i \frac{(-1)^{|\aaa|}}{4^{|\aaa|}}\qtx{and}
\frac{\sgn(\aaa,\bar\aaa+\lt\bar a^\rc_{1k}\rt,\aaa^\brc,\bar\aaa^\brc-\lt\bar a^\rc_{1k} \rt)}
{\sgn(\aaa,\bar\aaa,\aaa^\brc,\bar\aaa^\brc )}=(-1)^{m+a^\rc_{1k}},
\eeq
similar calculations  lead to
\begin{align}
 & \int e^{i\BPhi\cdot\BPhi'} {\vpsi} \cdot \f(\BPhi^\ot) 
\;D\BPhi  \label{eq-vec-int2} \\
& =-2i  \sum_{(\bar\aaa,\aaa)\in\Pp^n_{-1}} \bigg\{ \sum_{k=1}^{|\bar a^\rc_1|} 
(-1)^{m+\bar a^\rc_{1k}+k-1+|a^\rc_1|}
 D_{\bar\aaa+\lt\bar a^\rc_{1k}\rt, \aaa} \,T f_{\bar a^\rc_{1k}} (\bvp'^\ot) \bigg\} \sgn(\bar\aaa) \Psi'_{\bar\aaa,\aaa} \notag \\
& =-2i   \sum_{(\bar\aaa,\aaa)\in\Pp^n_{-1}} \bigg\{  \sum_{k=1}^{|\bar a^\rc_1|} 
(-1)^{\bar a^\rc_{1k}-k+|\bar a_1|} 
 D_{\bar\aaa+\lt\bar a^\rc_{1k}\rt, \aaa} \,T f_{\bar a^\rc_{1k}} (\bvp'^\ot) \bigg\}  \sgn(\bar\aaa) \Psi'_{\bar\aaa,\aaa} \notag\;.
\end{align}

We are now ready to derive the main identities we will need to prove Theorem~\ref{theo-main1}. The integrals \eqref{eq-vec-int} and
\eqref{eq-vec-int2} can be expressed using the matrix  operator $\T$ defined by
\begin{equation} \label{eq-def-sT}
 \T \f
:= 2 \bpart\,T \f\qtx{for} \f=(f_1,\ldots,f_m)^\top\in [\Ss_n(\Sym^+(m))]^m\;.
\end{equation}
Combining
\eqref{eq-DT-FD}, \eqref{eq-DD=d}, \eqref{eq-dT-F} and \eqref{eq-T2=1} we see that $\T$ is an involution,
\begin{equation}
 \T^2\f= 4 \bpart T \bpart T \f = 4 (-2)^{mn-2} \Ff \delta^{n-1} \bslp \bpart T \f =
(-2)^{mn} \Ff \dlt^n T \f = T^2\f= \f\;.
\end{equation}

The following result is crucial for this article. It is the key observation that allows the extension of the results in 
\cite{K8} to the Bethe strip.
\begin{theorem} \label{th-DT}
Let $\f=(f_1, f_2 \ldots, f_m)^\top \in [S_n(\Sym^+(m))]^m$. Then 
\begin{align} \label{eq-DT}
 {\vbpsi'} \cdot \T\f (\BPhi'^\ot) &= 
i\int e^{i\BPhi\cdot\BPhi'} {\vbpsi} \cdot \f(\BPhi^\ot)
D\BPhi =
-i\int e^{-i\BPhi\cdot\BPhi'} {\vbpsi} \cdot \f(\BPhi^\ot)
D\BPhi ,
\end{align}
 and
\begin{align} \label{eq-1DT}
{\vpsi'} \cdot \T\f (\BPhi'^\ot) &= 
 i \int e^{i\BPhi\cdot\BPhi'} {\vpsi} \cdot \f(\BPhi^\ot)
D\BPhi =
-i \int e^{-i\BPhi\cdot\BPhi'} {\vpsi} \cdot \f(\BPhi^\ot)
D\BPhi .
\end{align}
\end{theorem}

\begin{proof}
The second equalities in \eqref{eq-DT} and \eqref{eq-1DT} follow from a simple change of variables.
Using \eqref{eq-vexp-1} we get
\begin{align} \label{eq-DT1}
& {\vbpsi'} \cdot \bpart T \f(\BPhi'^\ot) \\ \notag
& \qquad = \sum_{(\bar\aaa,\aaa)\in\Pp^n_1}\set{ \sum_{k=1}^{|\bar a_1|} \sum_{k'=1}^m (-1)^{k-1} D_{\bar\aaa-\lt\bar a_{1k}\rt,\aaa} 
D_{\{a_{1k}\},\{k'\}} Tf_{k'}(\bvp'^\ot)}\sgn(\aaa) \Psi'_{\bar\aaa,\aaa}\;.
\end{align}
First consider the case $k'=a^\rc_{1j}$. There are $a^\rc_{1j}-1$ numbers smaller than $a^\rc_{1j}$ in $\I$, $j-1$ of them are in the set
$a^\rc_1$ and hence $a^\rc_{1j}-j$ of them are in the set $a_1$. Therefore, $a^\rc_{1j}$ is the
$(a^\rc_{1j}-j+1)$-th smallest element of the set $a_1\cup\{a^\rc_{1j}\}$.
A column expansion of $D_{\bar a_1, a_1\cup\{a^\rc_{1j}\}}$, the determinant of
$\bpart_{\bar a_1, a_1\cup\{a^\rc_{1j}\}}$, leads to
\begin{equation}
D_{\bar a_1, a_1\cup\{a^\rc_{1j}\}}=
\sum_{k=1}^{|\bar a_1|} (-1)^{k+a^\rc_{1j}-j+1} D_{\bar a_1\setminus\{\bar a_{1k}\}, a_1} 
D_{\{\bar a_{1k}\},\{a^\rc_{1j}\}}\; ,
\end{equation}
implying
\begin{equation} \label{eq-DT2}
\sum_{k=1}^{|\bar a_1|} (-1)^{k-1} D_{\bar\aaa-\lt\bar a_{1k}\rt,\aaa} 
D_{\{ a_{1k}\},\{a^\rc_{1j}\}} 
= (-1)^{a^\rc_{1j}-j} D_{\bar\aaa,\aaa+\lt a^\rc_{1j} \rt} \;.
\end{equation}
Similarly, for $k'\in a_1$  we can also interpret the sum over $k$ as the expansion of a determinant. However, in this case
the corresponding matrix has two identical rows, therefore
\begin{equation} \label{eq-DT3}
 \sum_{k=1}^{|\bar a_1|} (-1)^{k-1} D_{\bar\aaa-\lt\bar a_{1k}\rt,\aaa} 
D_{\{ a_{1k}\},\{k'\}} Tf_{k'}(\bvp'^\ot)=0\qtx{for} k'\in a_1\;.
\end{equation}
Now equations \eqref{eq-DT1}, \eqref{eq-DT2} and \eqref{eq-DT3} lead to
\begin{equation} \label{eq-DT4}
\vbpsi' \cdot \bpart T \f (\BPhi'^\ot) =
\sum_{(\bar\aaa,\aaa)\in\Pp^n_1} \set{\sum_{j=1}^{|a^\rc_1|} 
(-1)^{a^\rc_{1j}-j} D_{\bar\aaa,\aaa+\lt a^\rc_{1j}\rt}
Tf_{a^\rc_{1j}}(\bvp'^\ot)}\sgn(\aaa) \Psi'_{\bar\aaa,\aaa}\,  ,
\end{equation}
which combined with \eqref{eq-vec-int} proves \eqref{eq-DT}.

For the second equation one starts from \eqref{eq-vexp-2}; similar calculations  lead to
\begin{align}
&{\vpsi'} \cdot \bpart T \f (\BPhi'^\ot) \\
& \qquad =
\sum_{(\bar\aaa,\aaa)\in\Pp^n_{-1}} \set{ \sum_{j=1}^{|a^\rc_1|} 
(-1)^{a^\rc_{1j}-j+|\bar a_1|} D_{\bar\aaa+\lt\bar a^\rc_{1j}\rt,\aaa}
Tf_{\bar a^\rc_{1j}}(\bvp'^\ot)} \sgn(\aaa) \Psi'_{\bar\aaa,\aaa}\;.  \notag
\end{align}
Combining this with \eqref{eq-vec-int2} yields \eqref{eq-1DT}.
\end{proof}

Next we introduce a Hilbert space on which   $\T$ is a unitary operator.
In view of \eqref{eq-vexp-1}, we define differential operators  on
$ [C^\infty_n(\Sym^+(m))]^m$ by 
\begin{equation}\label{eq-def-DDaa}
 \DD_{\bar\aaa,\aaa} \f :=  \sgn(\aaa)
\sum_{k=1}^{|\bar a_1|} (-1)^{k-1} D_{\bar\aaa-\lt\bar a_{1k}\rt,\aaa} f_{\bar a_{1k}}\qtx{for}
(\bar\aaa,\aaa)\in\Pp_1^n\;,
\end{equation}
where $ \f=(f_1,\ldots,f_m)^\top \in [C^\infty_n(\Sym^+(m))]^m$.
$\DD_{\bar\aaa,\aaa}$ may be considered as a row-vector of differential operators. 
Equations \eqref{eq-vexp-1} and \eqref{eq-vexp-2} can be written as
\begin{equation} \label{eq-vexp-DD}
 \vbpsi \cdot \f(\BPhi^\ot) = \sum_{(\bar\aaa,\aaa)\in\Pp_1^n} \DD_{\bar\aaa,\aaa} \f(\bvp^\ot) \Psi_{\bar\aaa,\aaa}
\end{equation}
and
\begin{equation} \label{eq-vexp-DD2}
{\vpsi} \cdot \f(\BPhi^\ot)=\sum_{(\bar\aaa,\aaa)\in\Pp_{-1}^n} (-1)^{|\bar a_1|}\DD_{\aaa,\bar\aaa} \f(\bvp^\ot) \Psi_{\bar\aaa,\aaa}\;.
\end{equation}
Combining Theorem~\ref{th-DT} with \eqref{eq-vec-int} and \eqref{eq-vexp-DD}, we  obtain 
\begin{equation}\label{eq-TT-parts}
 \DD_{\bar\aaa,\aaa} (\T \f) =  \frac{2^{mn+1}}{4^{|\bar\aaa|}} \frac{\sgn(\aaa)\sgn(\III)(-1)^{mn}}
{\sgn(\bar\aaa)\sgn(\aaa,\bar\aaa,\aaa^\brc,\bar\aaa^\brc)}\;
\Ff (\DD_{\aaa^\brc,\bar\aaa^\brc} \f)\qtx{for} (\bar\aaa,\aaa)\in\Pp^n_1\;.
\end{equation}
This leads us to define the norm
\begin{equation}\label{defHh}
 \| \f \|^2_{\HH} :=
\sum_{(\bar\aaa,\aaa)\in\Pp_1^n} \left\| 2^{|\aaa|} \DD_{\bar\aaa,\aaa} \f(\bvp^\ot) \right\|^2_{L^2(\RR^{m\times 2n};d^{2mn} \bvp)}\;,
\end{equation}
and let $\HH$ be the Hilbert space  completion of $[\Ss_n(\Sym^+(m))]^m$ with respect to  the norm $\|\cdot\|_\HH$.
By \eqref{eq-TT-parts},  $\T$ extends to a unitary operator on $\HH$.
Moreover, the expressions $\vbpsi \cdot\f$ and ${\vpsi} \cdot \f$ can be extended to $\f\in\HH$ and the
equations \eqref{eq-DT} and \eqref{eq-1DT} remain valid.

We also introduce the 
 Hilbert space tensor product $\KK:=\HH\otimes\HH$.
For $\f, \g \in \HH$ the tensor product
\hbox{$\f (\bvp_+^\ot)\otimes \g(\bvp_-^\ot)$} can be identified with the matrix valued function given by the matrix product 
$\f(\bvp_+^\ot)\, [\g(\bvp_-^\ot)]^\top$.
With this identification, the norm for an $m\times m$ matrix valued function $\bF(\bvp_+^\ot,\bvp_-^\ot)\in \KK$ is given by
\begin{equation} \label{eq-norm-KK}
 \| \bF \|^2_{\KK} :=
\sum_{\substack{(\bar\aaa,\aaa)\in\Pp_1^n\\(\bar\bbb,\bbb)\in\Pp_1^n}} 
\left\| 2^{|\aaa|+|\bbb|} 
\DD^{(-)}_{\bar\bbb,\bbb}\left[\DD^{(+)}_{\bar\aaa,\aaa} \bF(\bvp_+^\ot,\bvp^\ot_-)\right]^\top  
\right\|^2_{L^2(\RR^{m\times 4n};d^{4mn} (\bvp_+,\bvp_-))}\; ,
\end{equation}
where   $\DD_{\bar\aaa,\aaa}^{(\pm)}$ denotes   the operator
$\DD_{\bar\aaa,\aaa}$ acting with respect to  $\bvp^\ot_\pm$.
To obtain \eqref{eq-norm-KK}, note that $\DD^{(\pm)}_{\bar\aaa,\aaa}$ are $1\times m$ row-vectors of differential operators and hence
 $\DD^{(+)}_{\bar\aaa,\aaa} \f(\bvp_+^\ot)= \left[\DD^{(+)}_{\bar\aaa,\aaa} \f(\bvp_+^\ot)\right]^\top$, since it is a $1 \times 1$ matrix,  which leads to
\begin{equation} 
 \left[\DD^{(+)}_{\bar\aaa,\aaa} \f(\bvp_+^\ot)\right]\left[\DD^{(-)}_{\bar\bbb,\bbb} \g(\bvp_-^\ot)\right]
=\DD^{(-)}_{\bar\bbb,\bbb} \left[\DD^{(+)}_{\bar\aaa,\aaa} \f(\bvp_+^\ot) \g^\top(\bvp_-^\ot))\right]^\top\;.
\end{equation}
Together with \eqref{eq-vexp-DD} and \eqref{eq-vexp-DD2} this calculation also implies
\begin{equation} \label{eq-PFP}
 \vbpsi_+ \cdot \bF(\BPhi_+^\ot, \BPhi_-^\ot) \vbpsi_- =
\sum_{\substack{(\bar\aaa,\aaa)\in\Pp_1^n\\(\bar\bbb,\bbb)\in\Pp_1^n}} 
\DD^{(-)}_{\bar\bbb,\bbb}\left[\DD^{(+)}_{\bar\aaa,\aaa} \bF(\bvp_+^\ot,\bvp^\ot_-)\right]^\top  
\Psi_{+,\bar\aaa,\aaa}\, \Psi_{-,\bar\bbb,\bbb}\; 
\end{equation}
and
\begin{align}\label{eq-PFP2}
 &\vpsi_+ \cdot \bF(\BPhi_+^\ot, \BPhi_-^\ot) \vpsi_- = \\
& \qquad \qquad\qquad \qquad \sum_{\substack{(\bar\aaa,\aaa)\in\Pp_{-1}^n\\(\bar\bbb,\bbb)\in\Pp_{-1}^n}}\!\!\! (-1)^{|\bar a_1| + |\bar b_1|}
\DD^{(-)}_{\bbb,\bar\bbb}\left[\DD^{(+)}_{\aaa,\bar\aaa} \bF(\bvp_+^\ot,\bvp^\ot_-)\right]^\top  
\Psi_{+,\bar\aaa,\aaa}\, \Psi_{-,\bar\bbb,\bbb}\; ,\notag
\end{align}
where $\Psi_{+,\bar\aaa,\aaa}$ and $\Psi_{-,\bar\bbb,\bbb}$ are defined analogously to $\Psi_{\bar\aaa,\aaa}$
using the Grassmann entries of $\BPhi_+$ and $\BPhi_-$, respectively. An important operator on $\KK$ is the tensor operator $\wT:=\T\otimes\T$.
Theorem~\ref{th-DT} implies the following.
\begin{coro} \label{cor-DT}
Let  $\bF(\BPhi_+^\ot,\BPhi_-^\ot)\in \KK$.  Then
\begin{align}
 &\vbpsi'_+ \cdot \wT \bF (\BPhi'^\ot_+,\BPhi'^\ot_-) {\vbpsi'}_- \\ \notag
& \qquad\qquad = \int e^{\pm i(\BPhi_+ \cdot \BPhi'_+ -  \BPhi_- \cdot \BPhi'_-)} \vbpsi_+ \cdot \bF (\BPhi_+^\ot,\BPhi_-^\ot)
{\vbpsi}_-
D\BPhi_+ D\BPhi_-
\end{align}
and
\begin{align}
 &{\vpsi'_+} \cdot \wT\bF (\BPhi'^\ot_+,\BPhi'^\ot_-) {\vpsi'}_- \\ \notag
& \qquad\qquad = \int e^{\pm i(\BPhi_+ \cdot \BPhi'_+ - \BPhi_- \cdot \BPhi'_-)} \vpsi_+ \cdot \bF (\BPhi_+^\ot,\BPhi_-^\ot)
{\vpsi}_-
D\BPhi_+ D\BPhi_- \;.
\end{align}
\end{coro}

We recall that  $\Hh^m=\bigoplus_{k=1}^m \Hh$ and $\Kk^{m\times m}\cong \Kk^{m^2}$  are Hilbert spaces with the norms
\begin{align}
\|\f\|^2_{\Hh^m}&=\sum_{k=1}^m \hn f_k\hn_2^2 \qtx{for} \f=(f_1,\ldots,f_m) \in \Hh^m, \\
 \|\bF\|^2_{\Kk^{m\times m}}& =\sum_{j,k=1}^m \hnn F_{jk}\hnn_2^2  \qtx{for} \bF=(F_{jk})\in\Kk^{m\times m}.
\end{align}
We let $\wt$ act on $\Kk^{m\times m}$ by acting on all entries.
The relations between $\Hh^m$ and $\HH$ and between $\Kk^{m\times m}$ and $\KK$ will play a crucial role.

\begin{prop}\label{prop-K-K} $\quad$

\begin{enumerate}[\upshape (i)]
\item  $\Hh^m$ is a subset of $\HH$  and  the canonical injection  $ \Hh^m \mapsto\HH$
is continuous with respect to  the norms of $\Hh^m$ and $\HH$.

\item  $\Kk^{m\times m}$ is a subset of $\KK$ and the canonical injection $\Kk^{m\times m} \to \KK$ is continuous
with respect to  the norms of $\Kk^{m\times m}$ and $\KK$.
\item  The matrix differential operator $\bpart$ acting on $[\Ss_n(\Sym^+(m))]^m$ extends to a continuous operator
from $\Hh^m$ to $\HH$,  and we have
\begin{equation} \label{eq-bpart-TT}
2\bpart\, \f = \T \,T\, \f  \qtx{for all} \f \in \Hh^m .
\end{equation}

\item The operators $\bpart\otimes\one$, $\one\otimes\bpart$ and $\bpart\otimes\bpart$ are continuous
from $\Kk^{m\times m}$ to $\KK$. They are given by
\begin{eqnarray} \label{eq-bp+}
 (\bpart\otimes\one) \bF(\BPhi^\ot_+,\BPhi^\ot_-)&=&\bpart_+ \bF(\BPhi^\ot_+,\BPhi_-^\ot), \\
\label{eq-bp-}
(\one\otimes\bpart) \bF(\BPhi^\ot_+,\BPhi^\ot_-)&=&\left[\bpart_- \bF^\top(\BPhi^\ot_+,\BPhi_-^\ot)\right]^\top ,\\
\label{eq-bp+bp-}
(\bpart\otimes\bpart) \bF(\BPhi^\ot_+,\BPhi^\ot_-)&=&\left\{\bpart_-\left[\bpart_+ \bF(\BPhi^\ot_+,\BPhi_-^\ot)\right]^\top \right\}^\top ,
\end{eqnarray}
where $\bpart_\pm$ is the matrix of differential operators $\bpart$ with respect to  $\BPhi_\pm$. (The products are  matrix products.)

\item  We have 
\begin{equation}
\wT \bF  = 4 [\bpart_- (\bpart_+ \wt \bF)^\top ]^\top \qtx{for all} \bF\in\KK.
\end{equation}

\item
Given  $g(\bvp^\ot)\in\widetilde{\Hh}^{(0)}_\infty$,  the   multiplication operator $M(g)$,  defined by
\beq 
 M(g)\f(\bvp^\ot)=g(\bvp^\ot)\f(\bvp^\ot)\qtx{for} \f \in [\Ss_n(\Sym^+(m))]^m,
\eeq 
 extends to a  bounded operator on $\HH$.
The map $g \in \Hh^{(0)}_\infty\mapsto M(g) \in B(\HH)$ is continuous, and hence extends to $\Hh_\infty$.
Moreover,  for $(\bar\aaa,\aaa)\in\Pp_1^n$ we have
\begin{equation} \label{eq-Leibn1}
 \DD_{\bar\aaa,\aaa} (g\f)=
\sum_{\substack{(\bar\bbb,\bbb)\in\Pp^n, (\bar\bbb',\bbb')\in\Pp_1^n \\ (\bar\bbb,\bbb)+(\bar\bbb',\bbb')=(\bar\aaa,\aaa)}}
\sgn(\bbb)\sgn(\bar\bbb,\bbb,\bar\bbb',\bbb') D_{\bar\bbb,\bbb}\, g \;\DD_{\bar\bbb',\bbb'} \f
\end{equation}

\item Given  $G(\bvp^\ot_+,\bvp_-^\ot) \in \Kk_\infty^{(0)}$, the multiplication operator
$M(G)$, defined by
\beq
M(G)\bF (\bvp^\ot_+,\bvp_-^\ot)=G(\bvp^\ot_+,\bvp_-)\bF(\bvp_+^\ot,\bvp_-^\ot)\qtx{for}
\eeq
for $ \bF \in [\Ss_n(\Sym^+(m))]^m \otimes [\Ss_n(\Sym^+(m))]^m$,  extends to a  bounded operator 
 on $\KK$.  The map $G\in \Kk_\infty^{(0)} \mapsto M(G) \in B(\KK)$ is continuous, and hence extends to $\Kk_\infty$.

\end{enumerate}
\end{prop}

\begin{proof}
 (i) is a simple consequence of the definitions of the norm.  (ii) follows from (i) since $\Hh^m \otimes \Hh^m \cong \Kk^{m\times m}$.

To get (iii) note that for  $ \f \in [\Ss_n(\Sym^+(m))]^m$  \eqref{eq-bpart-TT} follows from \eqref{eq-T2=1} and \eqref{eq-def-sT}. Since $T$ is unitary on
$\Hh^m$, $\Hh^m$ is continuously embedded in $\HH$ by (i),  and $\T$ is unitary on $\HH$,
we conclude that the operator $\frac12 \T T$ defines a continuous linear map from $\Hh^m$ to $\HH$ which extends the map $\f\in  [\Ss_n(\Sym^+(m))]^m\mapsto \bpart \f \in  [\Ss_n(\Sym^+(m))]^m$.

The continuity in (iv) follows from (iii). 
For the second and third equation note that $\bpart=\bpart^\top$ and hence
\begin{align}
 (\one\otimes\bpart)(\f(\bvp^\ot_+)\otimes\g(\bvp_-^\ot)) &=
\f(\bvp^\ot_+) \left[\bpart_- \g(\bvp_-^\ot)\right]^\top 
= \left[\bpart_- \g(\bvp_-^\ot) \f^\top(\bvp^\ot_+)\right]^\top \notag \\
&= \left[\bpart_- \left[\f(\bvp^\ot_+)\g^\top(\bvp_-^\ot) \right]^\top\right]^\top.
\end{align}

(v) follows from (iv) and \eqref{eq-def-sT}.
To prove  (vi) note that $\vbpsi \cdot M(g)\f(\BPhi^\ot)=g(\BPhi^\ot) \vbpsi \cdot \f(\BPhi^\ot)$.
By \eqref{eq-vexp-DD} this implies
\eqref{eq-Leibn1} which leads to
$\|M(g)\f \|_{\HH}\leq C \hn g \hn_\infty \|\bF\|_{\HH} $ for a constant $C$ only depending on $m$ and $n$ which are fixed.
(vii) is proved  similarly  to (vi) considering $\vbpsi_+ \cdot M(G)\bF(\BPhi_+^\ot,\BPhi_-^\ot) {\vbpsi}_-$.
\end{proof}


\section{Averages of the matrix Green's function \label{sec-avgreen}}

We have all the main supersymmetric identities by now. So let us consider the random Hamiltonian
$H_\lb$  introduced in \eqref{Hlambda} and \eqref{Hlambda2}.
Recall that we fixed some arbitrary site in $\B$ which we called the origin and denoted by $0$.  
Given two nearest neighbors 
sites $x,y \in \B$, we will denote by ${\B}^{(x|y)}$ the lattice  obtained by removing from $\B$
the branch emanating from $x$ that passes through $y$; if we do not specify which 
branch was removed we will simply write ${\B}^{(x)}$.  Each vertex in  ${\B}^{(x)}$ has degree 
$K +1$, with the single exception of $x$ which has degree $K$. 
Given $\La \subset \B$, we will use $\,H_{\lb, \La}$ to denote the operator $\,H_{\lb}$  
restricted to $\ell^2 (\Lambda,\CC^m)$ with Dirichlet boundary conditions.
The matrix Green's function corresponding to $\,H_{\lb, \La}$ will be denoted by  	
\begin{equation}
G_{\lb, \La}\, (x,y;z)\;\;=\;\;\left[\left\langle {x,j|(H_{\lb, \La} -z)^{-1}|y,k} \right\rangle\right]_{j,k\in\{1,\ldots,m\}} 
\end{equation}
for $x,y \in {\La}$, 
and $z = E + i\eta$ with $E \in \RR$, $\eta> 0$.

Important choices of $\Lambda\subset \B$ will be  the sets $\B_\ell$, denoting all sites $y\in \B$ 
with distance $|y|=d(0,y)\leq \ell$, and
$\B^{(x|y)}_\ell$ denoting all sites $x' \in \B^{(x|y)}$ with $d(x,x')\leq \ell$.
We will use the Green's matrix at the origin very often, therefore let us define
\beq
G_\lb(z):=G_\lb(0,0;z)\;.
\eeq
For special choices of $\Lambda$ let us also introduce the following notation:
\begin{equation}
\begin{array}{lclclcl}
H_{\lb,\ell}            &:=&  H_{\lb,\B_\ell}      &\qquad& 
G_{\lb,\ell}(z)         &:=&  G_{\lambda,\B_\ell}(0,0;z)  \\
H_{\lb}^{(x|y)}      &:=&  H_{\lb, {\B}^{(x|y)}} & &
G_{\lb}^{(x|y)}(z)   &:=&  G_{\lb,{\B}^{(x|y)}}(x,x;z) \\
H_{\lb,\ell}^{(x|y)}    &:=&  H_{\lb, {\B}^{(x|y)}_\ell} & &
G_{\lb,\ell}^{(x|y)}(z) &:=&  G_{\lb,{\B}^{(x|y)}_\ell}(x,x;z) \\
H_{\lb}^{(x)}        &:=&  H_{\lb, {\B}^{(x)}} & &
G_{\lb}^{(x)}(z)     &:=&  G_{\lb,{\B}^{(x)}}(x,x;z) \\
\end{array}
\end{equation}
Similarly to \cite[Prop. 1.2]{AK}, we have
\begin{equation}\label{eq-lim-l}
 \lim_{\ell\to\infty} G_{\lb,\B_\ell}(x,y;z)\;=\;G_\lb(x,y;z)\; \; \text{and}\;\;
\lim_{\ell\to\infty} G_{\lambda,\B_{x,\ell}}(x,y;z)\;=\;
G_\lb(x,y;z)\;.
\end{equation}

To each site $x\in\B$ we assign supermatrices 
$\BPhi_x$, $\BPhi_{x,+}$,  and $\BPhi_{x,-}$, which are all independent, i.e., all different Grassmann variables are independent.
We will also use the independent supermatrices $\BPhi,\BPhi',\BPhi_+,\BPhi_-, \BPhi'_+$ and $\BPhi'_-$.
Furthermore we may use notations like $\BPhi_x=(\bvp_x,\BPsi_x)$ where $\bvp_x$ is a variable varying in
$\RR^{m\times 2n}$ and $\BPsi_x=((\overline{\psi}_x)_{k,\ell},(\psi_x)_{k,\ell})_{k,\ell}$.
Also $\vbpsi_x, \vbpsi_{x,\pm}$ and so on shall be defined analogously to \eqref{eq-def-vpsi}.

For each finite subset $\Lambda\subset\B$ we
set $D_\Lambda\BPhi\;=\;\prod_{x\in\Lambda} D\BPhi_x$, where $D\BPhi_x$ is defined 
as in \eqref{eq-def-DPhi}.
Let $B$ be an operator on $\ell^2(\B,\CC^m)$ and $B_\Lambda$ its restriction to $\ell^2(\Lambda,\CC^m)$ for a finite 
set $\Lambda\subset\B$. 
For $x,y\in\Lambda$ we define $\langle x|B_\Lambda|y \rangle$ 
to be the $m\times m$ matrix with entries $(\langle x,j|B_\Lambda|y,k\rangle)_{j,k}$.
Furthermore we define 
\begin{equation}
\langle \BPhi|B_\Lambda|\BPhi\rangle\;=\; \sum_{x,y\in\Lambda}\, \BPhi_x\,\cdot\;\langle x\,|\,B_\Lambda\,|\,y\,\rangle\,\BPhi_y\;.
\end{equation}
Now let $\im z>0$,  $\Lambda\subset\B$  finite, and $x,y\in\Lambda$. 
By the supersymmetric replica trick,  for any replica $s\in\{1,\ldots,n\}$ we have,  as  in \cite{B,E,K},
\begin{equation}\label{eq-G}
 \left[G_{\lambda,\Lambda}(x,y;z)\right]_{j,k}\;=\;
i\int (\psi_x)_{j,s} (\overline{\psi}_y)_{k,s}\,e^{-i\langle\BPhi|H_{\lambda,\Lambda}-z|\BPhi\rangle}\,D_{\Lambda}\BPhi\,.
\end{equation}

For some fixed $x\in\B$ we will denote by $x_0=0, x_1,\ldots,{x}_{|x|}=x$ the shortest path from $0$ to $x$, 
{\it i.e.}, $d(x_i,x_{i-1})=1$ and $x_i \neq x_j$ for $i\neq j$. 
We denote by $\B_{x,\ell}$ all sites in $\B$ whose distance from the path $x_0,\ldots,x_{|x|}$ is
at most $\ell+1$. If we let $N(x_i)$ be the set of neighbors of $x_i$ which are not on the path $x_0,\ldots,x_{|x|}$, then,
as a set,
\begin{equation}
 \label{eq-Bxl}
\B_{x,\ell}\;=\;\{x_0,x_1\ldots,x_{|x|}\}\,\cup\,\left[\bigcup_{i=0}^{|x|} \bigcup_{y\in N(x_i)} \B_\ell^{(y|x)} \right]\;,
\end{equation}
where the union is disjoint.
Note that for $|x|\geq1$ we have $|N(x_i)|=K-1$ for $i=1,\ldots,|x|-1$ and $|N(0)|=|N(x)|=K$.
If $|x|=0$, {\it i.e.} $x=0$, then $N(0)=K+1$.

Setting $\Lambda=\B_{x,\ell}$ in \eqref{eq-G}, noting that
$\big[{\vpsi}_0 \vbpsi^\top_x\big]_{j,k}=(\psi_0)_{j,1} (\overline{\psi}_x)_{k,1}$
and using the decomposition \eqref{eq-Bxl}
we obtain
\begin{equation} \label{eq-Gr-1}
 G_{\lb,\B_{x,\ell}}(0,x;z)\;=\;
i \int {\vpsi}_0 {\vbpsi}_x^\top
\prod_{j=0}^{|x|-1} e^{-i\BPhi_{x_j}\cdot\BPhi_{x_{j+1}}}\prod_{j=0}^{|x|} {\Theta_j}\,D_{\B_{x,\ell}} \BPhi
\end{equation}
where
\begin{equation} \label{eq-Xi}
 \Theta_j\;=\;e^{i\BPhi_{x_j} \cdot (z-A-\lambda V(x_j)) \BPhi_{x_j}}\;
e^{-i\sum_{y\in N(x_j)} \left[\BPhi_{x_j}\cdot\BPhi_y+
\langle\BPhi|H_{\lb,\ell}^{(y|x_j)}|\BPhi\rangle \right] }\;.
\end{equation}

In order to simplify this equation note that one obtains as in \cite[eq. (3.11)]{KS},
\begin{equation} \label{eq-Gr-2}
\int\, e^{-i\BPhi_{x_j}\cdot\BPhi_y- i\langle \BPhi|H^{(y|x_j)}_{\lambda,\ell-1} -z|\BPhi\rangle} 
\,D_{\B^{(y|x_j)}_{\ell}}\BPhi
\;=\;
e^{(i/4)\,\BPhi_{x_j} \cdot G^{(y|x_j)}_{\lambda,\ell}(z)\BPhi_{x_j}}\;.
\end{equation}
Plugging \eqref{eq-Gr-2} into \eqref{eq-Gr-1}, using \eqref{eq-lim-l}, and letting 
 $\ell\to \infty$, we get 
\begin{equation} \label{eq-Gr-3}
 G_\lb(0,x;z)\;=\;
i \int {\vpsi}_0 {\vbpsi}^\top_x \prod_{j=0}^{|x|-1} e^{-i\BPhi_{x_j}\cdot\BPhi_{x_{j+1}}}
\prod_{j=0}^{|x|} \Upsilon^{x,j}_{\lb,z}(\BPhi_{x_j}^\ot) \prod_{j=0}^{|x|} D\BPhi_{x_j}\;,
\end{equation}
where
\begin{equation} \label{eq-up}
 \Upsilon^{x,j}_{\lb,z}(\bvp^\ot)\;=\;
e^{i\Tr\left(\left[z-\lambda V(x_j)-A+\frac14 \sum_{y\in N(x_j)} G_\lb^{(y|x_j)}(z)  \right]\bvp^\ot\right) }\;.
\end{equation}
The dependence on $x$ results from the fact that $x$ determines the path $x_0=0,x_1,\ldots, x_{|x|}=x$.
Now we want to consider $|G_\lb(0,x;z)|^2$.
To improve the appearance of the following equations, we introduce the following notation:
\begin{equation}
 \begin{array}{lll} 
\wBPhi := (\BPhi_{+}, \BPhi_{-}),   &
 \wBPhi^\ot := (\BPhi_{+}^\ot, \BPhi_{-}^\ot), &
D\wBPhi := D\BPhi_{+} \,D\BPhi_{-},  \\
\wbvp := (\bvp_{+}, \bvp_{-}),   &
\wbvp^\ot := (\bvp_{+}^\ot, \bvp_{-}^\ot), \\
\wBPhi_x := (\BPhi_{x,+}, \BPhi_{x,-}),  &
\wBPhi_x^\ot := (\BPhi_{x,+}^\ot, \BPhi_{x,-}^\ot),&
 D\wBPhi_x := D\BPhi_{x,+} \,D\BPhi_{x,-},
 \end{array}
\end{equation}
\begin{equation}
 \wBPhi_x \cdot \wBPhi_y := \BPhi_{x,+} \cdot \BPhi_{y,+} \,-\, \BPhi_{x,-} \cdot \BPhi_{y,-}\;.
\end{equation}

From \eqref{eq-Gr-3} we obtain
\begin{align} \label{eq-GG}
 & |G_\lb(0,x;z)|^2 = G^*_\lb(0,x;z) G_\lb(0,x;z)\\ \notag
& \quad = - \int {\vbpsi}_{x,+} {\vpsi}^\top_{0,+} \Gamma^{x,r}_{\lb,z}(\wBPhi^\ot_{x}) 
\prod_{j=0}^{|x|-1} \left[ e^{i\wBPhi_{x_j}\cdot\wBPhi_{x_{j+1}}}
\Gamma^{x,j}_{\lb,z}(\wBPhi^\ot_{x_j})\right]
{\vpsi}_{0,-} \vbpsi^\top_{x,-}
\prod_{j=0}^{|x|} D\wBPhi_{x_j},
\end{align}
where
\begin{align} \label{eq-Gm}
 \Gamma^{x,j}_{\lb,z}(\wbvp^\ot) &=
 \overline{\Upsilon}^{x,j}_{\lb,z}(\bvp_+^\ot) \Upsilon^{x,j}_{\lb,z}(\bvp_-^\ot),
\end{align}
 with the bar denoting complex conjugation. The minus sign in \eq{eq-GG} comes from 
\begin{equation}
 \left[ \vpsi_{0,+}{\vbpsi}^\top_{x,+} \right]^\top
=  - \vbpsi_{x,+}{\vpsi}^\top_{0,+},
\end{equation}
a consequence of the  anti-commutation relations for Grassmann variables.

As in \cite{KS}, for $\lb \in \RR$, $E \in \RR$ and $\eta > 0$ let us introduce $\xi_{\lb,z} \in \Kk_\infty$ by
\begin{equation}
 \xi_{\lb,z} (\wbvp^\ot) =
\E \left(\exp{ \left\{\frac{i}{4} \Tr\left ( G_{\lb}^{(0)}(z) \,\bvp_+^\ot\, - \,
\left[{ G_{\lb}^{(0)}(z) }\right]^*\,\bvp_-^\ot\right)\right\}}\right)    \label{xia} \; ,
\end{equation}
and the operator $\Bb_{\lb,z}$  by
\begin{equation}
{\Bb}_{\lb,z} = 
M(\e^{i \Tr((z-A)\bvp_+^\ot\,-\,(\bar z -A)\bvp_-^\ot]} h(\lb (\bvp_+^\ot-\bvp_-^\ot))  )\;,
\end{equation}
where
$M(g(\wbvp^\ot))$ denotes multiplication by the function $g(\wbvp^\ot) $.
The fact that $\xi_{\lb,z}\in\Kk_\infty$ is a continuous family of elements is shown in \cite{KS}.
$\Bb_{\lb,z}$ and $M(g)$ act on matrix valued functions by acting on each entry.
Very important will be \cite[eq. (4.12) and Theorem~5.6]{KS} stating the following.
\begin{theorem}\label{theo-xi}
For $E\in I_{A,K}$ there exists $\lb_E>0$ and $\varepsilon_E>0$, such that the continuous map
\begin{equation}
 (\lb,E',\eta) \in (-\lb_E,\lb_E)\times(E-\varepsilon_E, E+\varepsilon_E)\times(0,\infty) \mapsto
\xi_{\lb,E'+i\eta} \in \Kk_\infty
\end{equation}
has a continuous extension to 
$(-\lb_E,\lb_E)\times(E-\varepsilon_E, E+\varepsilon_E)\times[0,\infty)$ satisfying the fix point equation
\begin{equation}
 \label{eq-fxp}
\xi_{\lb,z}=\wt \Bb_{\lb,z} \xi^{K}_{\lb,z}\;
\end{equation}
in $\Kk_\infty$.
\end{theorem}

We set
\beq
\Xi=\set{\bigcup_{E\in I_{A,K}}(-\lb_E,\lb_E)\times(E-\varepsilon_E, E+\varepsilon_E)\times[0,\infty)} \cup \set{\RR \times \RR \times (0,\infty)}\;,
\eeq
where   $\lb_E>0$ and $\varepsilon_E>0$ are as in the theorem, so we can extend
 $\xi_{\lb,E+i\eta}$  to a continuous function  on all of   $\Xi$, defining $\xi_{\lb,E}$ for $(\lambda,E,0)\in \Xi$.

We further define $\bxi_{\lb,E + i\eta}$ on $\Xi$ to be the diagonal $m\times m$ matrix with $\xi_{\lb,E + i\eta}$ on all diagonal entries, i.e.
\begin{equation}
 \bxi_{\lb,E + i\eta} := \xi_{\lb,E + i\eta} \,\one  .
\end{equation}
Note that 
the map
$(\lb,E,\eta)\mapsto \bxi_{\lb,E + i\eta}\in\Kk^{m\times m}\subset \KK$ is also continuous on $\Xi$.

\begin{prop}
We have
\begin{align} \label{eq-EGG}
& \E |G_\lb(0,x;z)|^2 = \\ \notag
& -\int {\vbpsi}_+  {\vpsi}^\top_+ \left\{
\Bb_{\lb,z}M(\xi_{\lb,z}^{K})\left[\wT\Bb_{\lb,z} M(\xi_{\lb,z}^{K-1})\right]^{|x|} \bxi_{\lb,z}) \right\} (\wBPhi^\ot)
{\vpsi}_- {\vbpsi}^\top_- \, D\wBPhi  .
\end{align}
\end{prop}

\begin{proof}
Let $x\neq 0$. $\set{\Gamma^{x,j}_{\lb,z}(\wBPhi^\ot)}_{j=0,1,\ldots,|x|}$ are independent Grassmann algebra-valued random variables, with
\begin{equation}
 \E (\Gamma^{x,j}_{\lb,z}(\wBPhi^\ot)) =
\begin{cases}
 \Bb_{\lb,z} \xi_{\lb,z}^K(\wBPhi^\ot) & \text{if} \quad j=0 \;\;\text{or}\;\, j=|x| \\
 \Bb_{\lb,z} \xi_{\lb,z}^{K-1}(\wBPhi^\ot) & \text{if} \quad 0<j<|x|
\end{cases}.
\end{equation}
Thus,  taking expectation in \eqref{eq-GG} and using the matrix equality
\beq
\xi_{\lb,z}(\wBPhi_0^\ot) {\vpsi}_{0,-} \vbpsi^\top_{x,-}=\xi_{\lb,z}(\wBPhi_0^\ot) \one {\vpsi}_{0,-} \vbpsi^\top_{x,-}= \bxi_{\lb,z}(\wBPhi_0^\ot) {\vpsi}_{0,-} \vbpsi^\top_{x,-},
\eeq
we get
\begin{align} \label{eq-EGG2}
&\E |G_\lb(0,x;z)|^2 = - \int {\vbpsi}_{x,+} {\vpsi}^\top_{0,+} \Bb_{\lb,z} \xi^K_{\lb,z}(\wBPhi_x^\ot) \\
&  \quad \quad \notag   \times 
\set{\prod_{j=0}^{r-1} \left[ e^{i\wBPhi_{x_{j+1}}\cdot\wBPhi_{x_j}}
\Bb_{\lb,z} \xi^{K-1}_{\lb,z}(\wBPhi^\ot_{x_j})\right] }
\bxi_{\lb,z}(\wBPhi_0^\ot)
{\vpsi}_{0,-} \vbpsi^\top_{x,-}
\prod_{j=0}^r D\wBPhi_{x_j} \;.
\end{align}
Using Corollary~\ref{cor-DT}, integration over $D\wBPhi_0=D\wBPhi_{x_0}$ gives
\begin{align}
& \E |G_\lb(0,x;z)|^2 =\\ \notag &
 - \int {\vbpsi}_{x,+} {\vpsi}^\top_{x_1,+} \Bb_{\lb,z} \xi^K_{\lb,z}(\wBPhi_x^\ot) 
\set{\prod_{j=1}^{r-1} \left[ e^{i\wBPhi_{x_{j+1}}\cdot\wBPhi_{x_j}}
\Bb_{\lb,z} \xi^{K-1}_{\lb,z}(\wBPhi^\ot_{x_j})\right]} \\
&\qquad \qquad \qquad \qquad  \times 
\left[ \wT \Bb_{\lb,z} M(\xi^{K-1}_{\lb,z}) \bxi_{\lb,z}\right](\wBPhi_{x_1}^\ot)
{\vpsi}_{x_1,-} \vbpsi^\top_{x,-}
\prod_{j=0}^r D\wBPhi_{x_j} \;. \notag
\end{align}
Repeated similar integrations  over $D\wBPhi_{x_j}$ for $j=1,2,\ldots,r-1$,  yields \eqref{eq-EGG} after  renaming 
$\wBPhi_x=\wBPhi_{x_{|x|}}$ as $\wBPhi$.

For the case $x=0$ note that $\E (\Gamma^{0,0}_{\lb,z}(\wBPhi^\ot)) = \Bb_{\lb,z} \xi_{\lb,z}^{K+1}(\wBPhi^\ot)$, which gives
\eqref{eq-EGG} also for $x=0$.
\end{proof}

To write the trace of $|G_\lb(0,x;z)|^2$ in a more compact way, let us introduce the following notations. 
First let us define the operator
\begin{equation}\label{eq-def-Vv}
 \Vv_{\lb,z} := \wT \,\Bb_{\lb,z} M(\xi_{\lb,z}^{K-1})\;.
\end{equation}
 Note that  $\Vv_{\lb,z}$ is a bounded linear operator on $\KK$  in view of Proposition~\ref{prop-K-K}~(vii).
For $\bF, \bF' \in \KK$ we define the bilinear forms
\begin{align} \notag 
\llangle \bF \,|\, \bF' \rrangle &:=
- \Tr\left\{ \int \bF^\top(\wBPhi^\ot) {\vbpsi}_+ {\vpsi}^\top_+\bF'(\wBPhi^\ot) {\vpsi}_- {\vbpsi}^\top_-
\,D\wBPhi
\right\} \\
&\;=  - \int \left[\vbpsi_+ \cdot \bF(\wBPhi^\ot) {\vbpsi}_-\right] \left[\vpsi_+ \cdot \bF'(\wBPhi^\ot) {\vpsi}_-\right]
\,D\wBPhi \label{eq-<>},
\end{align}

and
\begin{equation} \label{eq-<>lz}
 \llangle \bF \,|\, \bF' \rrangle_{\lb,z} :=
\llangle \Bb_{\lb,z} M(\xi_{\lb,z}^{K-1}) \bF \,|\, \bF' \rrangle
=\llangle \bF \,|\, \Bb_{\lb,z}M(\xi_{\lb,z}^{K-1}) \bF' \rrangle\;.
\end{equation}
For the second equation in \eqref{eq-<>}, note that for matrices $\bF, \bF'$ whose entries are
even elements of the Grassmann algebra $\Lambda(\BPsi_+,\BPsi_-)$, we have
\begin{align}
\Tr\left\{\bF^\top {\vbpsi}_+ {\vpsi}^\top_+\bF'(\wBPhi^\ot) {\vpsi}_- {\vbpsi}^\top_-\right\} & = 
- \Big[\vbpsi_-\cdot \bF^\top {\vbpsi}_+\Big] \Big[\vpsi_+ \cdot \bF' {\vpsi}_-\Big] \notag \\
& = \Big[\vbpsi_+ \cdot \bF {\vbpsi}_-\Big] \Big[\vpsi_+ \cdot \bF' {\vpsi}_-\Big].
\end{align}
The sign changes are caused by the anti-commutation relations of the Grassmann variables.
\begin{prop}\label{prop-GGVv} The following identities hold.
\begin{align}
& \llangle \bF \,|\, \wT\bF' \rrangle =
\llangle \wT\bF \,|\, \bF' \rrangle\;, \label{eq-<T>} \\
& \llangle \bF \,|\, \Vv_{\lb,z} \bF' \rrangle_{\lb,z} =
\llangle \Vv_{\lb,z} \bF \,|\, \bF' \rrangle_{\lb,z}\;,\label{eq-<Vv>} \\
& \E\left( \Tr (|G_\lb(0,x;z)|^2)\right) = \llangle \bxi_{\lb,z} \,|\, \Vv_{\lb,z}^{|x|} \bxi_{\lb,z} \rrangle_{\lb,z}.
\label{eq-GG-Vv}
\end{align}
\end{prop}

\begin{proof}
Using Corollary~\ref{cor-DT} and \eqref{eq-<>} we get
\begin{align}
\llangle \bF \,|\, \wT\bF' \rrangle  &= 
-\int 
\Big[\vbpsi_+ \cdot \bF(\wBPhi^\ot) {\vbpsi}_- \Big] \,e^{i\wBPhi\cdot\wBPhi'} \Big[
\vpsi'_+ \cdot \bF'(\wBPhi'^\ot) {\vpsi'}_-  \Big]
D\wBPhi' D\wBPhi
\notag \\
&=-\int 
\Big[ \vbpsi'_+ \cdot \wT \bF(\wBPhi'^\ot) {\vbpsi'}_- \Big]
\Big[ \vpsi'_+ \cdot \bF'(\wBPhi'^\ot) {\vpsi'}_- \Big]
D\wBPhi'
\notag \\
&= \llangle \wT\bF\,|\, \bF' \rrangle\;.
\end{align}
Equation \eqref{eq-<Vv>} is now a consequence of \eqref{eq-<>lz} and \eqref{eq-<T>},
equation \eqref{eq-GG-Vv} follows from \eqref{eq-EGG}, \eqref{eq-<>} and \eqref{eq-<>lz}.
\end{proof} 

Similarly to \cite{K8}, for $z=E+i\eta$ with $\eta >0$ we define $\bth_{\lb,z}\in\KK$ by
\begin{equation} \label{eq-def-th}
 \bth_{\lb,z} := -2(\bpart\otimes\one+\one\otimes\bpart) \bxi_{\lb,,z} =
-2(\bpart_+ + \bpart_-) \bxi_{\lb,z}\;.
\end{equation}

Using $\bxi_{\lb,z}=\bxi_{\lb,z}^\top$ and $\bpart_- \bxi_{\lb,z} = [\bpart_- \bxi_{\lb,z}]^\top$, the second equation 
follows from \eqref{eq-bp+} and \eqref{eq-bp-}.
Proposition~\ref{prop-K-K}~(iv) implies that the map $(\lb,E,\eta)\mapsto\bth_{\lb,E+i\eta}\in\KK$ is continuous on $\Xi$.

\begin{lemma} \label{lem-tx-tt}
For $\eta=\im z >0$ we have
\begin{align}
 \llangle \bth_{\lb,z} \,|\, \Vv^{|x|}_{\lb,z} \bxi_{\lb,z} \rrangle_{\lb,z} &= 
\E\left(\Tr\left(\Im_\lb^{(x'|x)}(z)|G_\lb(0,x;z)|^2 \right)\right) \notag \\
&=\E\left(\Tr\left( \left|G_\lb(0,x;z)\sqrt{\Im_\lb^{(x'|x)}(z)} \right|^2\right)\right) > 0 ,
\label{eq-th-Vv-xi}
\end{align}
and
\begin{align}
\llangle \bth_{\lb,z} \,|\, \Vv^{|x|}_{\lb,z} \bth_{\lb,z} \rrangle_{\lb,z} &= 
\E\left(\Tr\left(\Im_\lb^{(x'|x)}(z)G^*_\lb(0,x;z) \Im_\lb^{(0'|0)}(z) G_\lb(0,x;z) \right)\right) \notag \\
&=\E\left(\Tr\left( \left|\sqrt{\Im_\lb^{(x'|x)}(z)}G_\lb(0,x;z)\sqrt{\Im_\lb^{(0'|0)}(z)} \right|^2\right)\right) > 0 ,
\label{eq-th-Vv-th}
\end{align}
where $0'\in\B$ is a neighbor of $0\in \B$ and $x'\in \B$ is a neighbor of $x\in \B$, both not lying on the path from 0 to $x$.
\end{lemma}
\begin{proof}
As $H_\lb$ is a real operator, the Green's matrices $G_\lb^{(x|y)}(z)$ are symmetric, implying
$\left[G_\lb^{(x|y)}(z)\right]^*=\overline{G_\lb^{(x|y)}(z)}$, where the overline denotes  complex conjugation.
It follows that the imaginary parts 
$\Im_\lb^{(x|y)}(z):=\frac{1}{2i}\left(G_\lb^{(x|y)}(z)- \left[G_\lb^{(x|y)}(z)\right]^*\right)$ 
are {\em real}, {\em symmetric} matrices. More over, $\Im_\lb^{(x|y)}$ is positive 
if $\eta=\im z>0$.
Since
\begin{equation} \label{eq-Dexp}
\bpart e^{\Tr(M\bvp^\ot)}= M e^{\Tr(M\bvp^\ot)}
\end{equation}
for symmetric $m \times m$ matrices $M$, we obtain
\begin{equation}
\bth_{\lb,z}(\wbvp^\ot)= 
\E \left(\Im_\lb^{(0)}(z)\exp{ \left\{\frac{i}{4} \Tr\left ( G_{\lb}^{(0)}(z) \,\bvp_+^\ot\, - \,
\left[ G_{\lb}^{(0)}(z) \right]^*\,\bvp_-^\ot\right)\right\}}\right)    \label{eq-th} \;.
\end{equation}
In particular, $\bth_{\lb,z}$ is a symmetric matrix.
The equation \eqref{eq-th} implies
\begin{equation} \label{eq-EBxth}
 \Bb_{\lb,z} M(\xi_{\lb,z}^{K-1})\bth_{\lb,z}=
\E\left(\Im_\lb^{(x'|x)}(z) \Gamma^{x,|x|}_{\lb,z} \right)
=\E\left(\Im_\lb^{(0'|0)}(z) \Gamma^{x,0}_{\lb,z} \right)\;.
\end{equation}
Multiplying \eqref{eq-GG} by $\Im_\lb^{(x'|x)}(z)$ from the left, taking expectations and combining this with
Corollary~\ref{cor-DT}, equations
\eqref{eq-<>}, \eqref{eq-<>lz}, \eqref{eq-EBxth} and the fact that $\bth_{\lb,z}^\top=\bth_{\lb,z}$ 
gives the first equation in \eqref{eq-th-Vv-xi}.
To get the first equation in \eqref{eq-th-Vv-th} we multiply \eqref{eq-GG} by $\Im_\lb^{(x'|x)}(z)$ from the left, insert the
matrix $\Im_\lb^{(0'|0)}(z)$ between the matrices $\vbpsi_{x,+} {\vpsi}_{0,+}^\top$ and $ \vpsi_{0,-} {\vbpsi}^\top_{x,-}$,
and  take expectations.

The only thing left to prove are the inequalities in \eqref{eq-th-Vv-xi} and \eqref{eq-th-Vv-th}.
Since $\Im_\lb^{(0'|0)}(z)$ and $\Im_\lb^{(x'|x)}(z)$ are both invertible, both inequalities will follow
if we can show that the matrix $G_\lb(0,x;z)$ is not identically zero for almost all potentials.

Let $H_\lb^{(0\not -x_1)}=H_\lb^{(0|x_1)} \oplus H_\lb^{(x_1|0)}$, then
$H_\lb=H_\lb^{(0\not - x_1)}+\Gamma$ where 
\begin{equation}
\langle y,k | \Gamma | z,j\rangle= 
\frac12\left(\delta_{y,0} \delta_{j,k} \delta_{z,x_1} + \delta_{y,x_1}\delta_{j,k} \delta_{z,0}\right)\;.
\end{equation}
Using the resolvent identity, 
\begin{equation}
(H_\lb-z)^{-1}= 
(H_\lb^{(0 \not - x_1)}-z)^{-1}- (H_\lb^{(0\not - x_1)}-z)^{-1}\Gamma (H_\lb-z)^{-1}\;,
\end{equation}
and the fact that $\langle 0| (H_\lb^{(0 \not - x_1)}-z)^{-1}| x\rangle=0$, we obtain the matrix equation
\begin{equation}
 G_\lb(0,x;z)=-\tfrac12 G_\lb^{(0|x_1)}(0,0;z) G_{\lb}(x_1, x ;z)\;.
\end{equation}
Iterating this procedure gives
\begin{equation}
 G_\lb(0,x;z)=(-\tfrac12)^{|x|} \left[\prod_{j=1}^{|x|} G^{(x_{j-1}|x_j)}_\lb(z) \right] G_\lb(x,x;z)\;.
\end{equation}
For $\eta=\im z>0$ the imaginary parts of these matrix Green's functions on the right hand side are positive.
Therefore all these matrices are invertible and hence $G_\lb(0,x;z)$ is invertible and hence not zero (for all random potentials).
\end{proof}

\section{The proof of Theorem~\ref{theo-main1} \label{sec-proof}}

From now on the proof is completely analogous to \cite{K8}.
By \eqref{eq-GG-Vv} we obtain
\begin{align}
 J_{\lb}(z):=\sum_{x\in\B} 
 |x|^2 \E\left( \Tr (|G_\lb(0,x;z)|^2)\right) 
= \frac{K+1}{K} \sum_{r=1}^\infty r^2 \llangle \bxi_{\lb,z} | \Ww_{\lb,z}^{r} \bxi_{\lb,z} \rrangle_{\lb,z},
\end{align}
where $\Ww_{\lb,z}=K\Vv_{\lb,z}$ and $\im z=\eta>0$.

\begin{lemma}  Let $\lb \in \RR$, $z=E + i \eta$ with $E \in \RR\,$ and $ \eta >0$.
\begin{enumerate}[\upshape (i)]
\item For all $\bF,\bF' \in \KK$ 
\begin{equation}
 \llangle \bF | \Ww_{\lb,z} \bF'\rrangle_{\lb,z} =
\llangle \Ww_{\lb,z} \bF | \bF' \rrangle_{\lb,z}\;.  \la{sym}
\end{equation}
\item
$\Ww_{\lb,z}^2$ is a compact operator on $\KK$.
\item  We have 
\begin{equation}
\Ww_{\lb,z} \bth_{\lb,z}=
\bth_{\lb,z} - \tfrac{4 \eta}{K}\Ww_{\lb,z}  \bxi_{\lb,z}\;.  
 \la{theta}
\end{equation}
\item For any $r=0,1,2,\ldots$ we have
\begin{equation}
 \llangle \bxi_{\lb,z}\, |\,  \Ww_{\lb,z}^{r} \,  \bth_{\lb,z}\rrangle_{\lb,z} >
 \llangle \bxi_{\lb,z}\, |\,  \Ww_{\lb,z}^{r+1} \,  \bth_{\lb,z}\rrangle_{\lb,z} > 0 \la{pos1}
\end{equation}
and
\begin{equation}
 \llangle \bth_{\lb,z}\, |\,  \Ww_{\lb,z}^{r} \,  \bth_{\lb,z}\rrangle_{\lb,z}\;\;>
\llangle \bth_{\lb,z}\, |\,  \Ww_{\lb,z}^{r+1} \,  \bth_{\lb,z}\rrangle_{\lb,z}\;\;>\;\;0\;.  \la{pos2}
\end{equation}
 \end{enumerate}
\end{lemma}

\begin{proof} (i) follows from \eqref{eq-<Vv>}. (ii) is a consequence of ${\Bb}_{\lb,z} M(\xi^{K-1}_{\lb,z}) \wT
{\Bb}_{\lb,z}$ being a compact operator on $\KK$
for $\eta >0$, which can be shown analogously to \cite[Lemma~5.1~(i)]{KS}, using 
\eqref{eq-TT-parts} as well as the
the Leibniz rules
\eqref{eq-Leibn} and \eqref{eq-Leibn1}.

 To prove (iii), note first that by \eqref{eq-dT-F}, \eqref{eq-DT-FD} and \eqref{eq-fxp} we have
 \begin{align}
\bth_{\lb,z} &=-2(\bpart\otimes \one+\one\otimes\bpart )\,{\wt}{\Bb}_{\lb,z} \bxi_{\lb,z}^K \notag \\ 
 &=(-2)^{2mn-1} (\Ff\dlt^{n-1} \bslp\otimes\Ff\dlt^n + \Ff\dlt^n \otimes \Ff\dlt^{n-1}\bslp)\,\Bb_{\lb,z} \bxi_{\lb,z}^K \notag \\
&=(-2)^{2mn-1} (\Ff\dlt^{n-1}\bslp\otimes \Ff\dlt^{n-1}\bslp)(\one\otimes\bpart+\bpart\otimes\one){\Bb}_{\lb,z} \bxi_{\lb,z}^K \notag \\
&=-2 \wT[(\bpart_+ + \bpart_-) {\Bb}_{\lb,z} \bxi_{\lb,z}^K] \;. \label{eq-th-expand}
\end{align}
Using \eqref{eq-Dexp} leads to
\begin{align} \notag
& -2(\bpart_+ + \bpart_- )
\left(e^{i\Tr\left((z-A)\bvp_+^\ot-(\bar z -A)\bvp_-^\ot \right)} h(\lb (\bvp_+^\ot-\bvp_- ^\ot))  \right) \\
& \qquad \qquad =
4\eta \left(e^{i\Tr\left((z-A)\bvp_+^\ot-(\bar z -A)\bvp_-^\ot \right)} h(\lb (\bvp_+^\ot-\bvp_- ^\ot))  \right)\;,
\end{align}
which combined with \eqref{eq-th-expand} gives
\begin{equation}
\bth_{\lb,z}=
-2 \wT[(\bpart_+ + \bpart_-) {\Bb}_{\lb,z} \bxi_{\lb,z}^K]
=
\Ww_{\lb,z} \bth_{\lb,z} + \tfrac{4 \eta}{K} \Ww_{\lb,z}  \bxi_{\lb,z}\;.
\end{equation}

The second inequalities in \eqref{pos1} and \eqref{pos2} follow from \eqref{eq-th-Vv-xi} and \eqref{eq-th-Vv-th}.
Using   \eqref{theta}, we have
\begin{align} \notag
 \llangle \bxi_{\lb,z}\, |\,  \Ww_{\lb,z}^{r+1} \,  \bth_{\lb,z}\rrangle_{\lb,z} &=
 \llangle \bxi_{\lb,z}\, |\,  \Ww_{\lb,z}^{r} \,  \bth_{\lb,z}\rrangle_{\lb,z}\;-\;
{\textstyle\frac{4 \eta}{K}} \llangle \bxi_{\lb,z}\, |\,  \Ww_{\lb,z}^{r+1} \,  \bxi_{\lb,z}\rrangle_{\lb,z} \\
& <\;\llangle \bxi_{\lb,z}\, |\,  \Ww_{\lb,z}^{r} \,  \bth_{\lb,z}\rrangle_{\lb,z} \;,
\end{align}
since  $\llangle \bxi_{\lb,z}\, |\,  \Ww_{\lb,z}^{r+1} \,  \bxi_{\lb,z}\rrangle_{\lb,z}>0$ by \eqref{eq-GG-Vv}.
Similarly,
\begin{align} \notag
 \llangle \bth_{\lb,z}\, |\,  \Ww_{\lb,z}^{r+1} \,  \bth_{\lb,z}\rrangle_{\lb,z} &=
 \llangle \bth_{\lb,z}\, |\,  \Ww_{\lb,z}^{r} \,  \bth_{\lb,z}\rrangle_{\lb,z}\;-\;
{\textstyle\frac{4 \eta}{K}} 
\llangle \bth_{\lb,z}\, |\,  \Ww_{\lb,z}^{r+1} \,  \bxi_{\lb,z}\rrangle_{\lb,z} \\
&<\llangle \bth_{\lb,z}\, |\,  \Ww_{\lb,z}^{r} \,  \bth_{\lb,z}\rrangle_{\lb,z} \;,
\end{align}
 since 
$ \llangle \bth_{\lb,z}\, |\,  \Ww_{\lb,z}^{r+1} \,  \bxi_{\lb,z}\rrangle_{\lb,z}>0$ by \eqref{eq-th-Vv-xi}.  
Thus (iv) is proven. 
\end{proof}

This lemma is just the generalization of \cite[Lemma~4.1]{K8} to the Bethe strip. Thus, from this point on we can use the exact 
same arguments as in \cite[ Lemmas~4.2, 4.3 and 4.4]{K8} to finally obtain
\begin{align} \label{eq-J-eta3}
 J_\lb(E+i\eta) \,\geq\, & \frac{K+1}{4\eta} \llangle \bxi_{\lb,z} \,|\, \bth_{\lb,z} \rrangle_{\lb,z} 
+ \frac{3K(K+1)}{16\eta^2} \llangle \bth_{\lb,z} \,|\, \bth_{\lb,z} \rrangle_{\lb,z}  \\
& \qquad \quad + \frac{K^2(K+1)}{64\eta^3} \frac{\left(\llangle \bth_{\lb,z} \,|\, \bth_{\lb,z} \rrangle_{\lb,z}-
\frac{4\eta}{K}\llangle\bxi_{\lb,z} \,|\, \bth_{\lb,z} \rrangle_{\lb,z}\right)^2}
{\llangle \bxi_{\lb,z} \,|\, \bth_{\lb,z} \rrangle_{\lb,z}} \; \notag
\end{align}
for $\lb\in\RR,\; z=E+i\eta$ with $E\in\RR$ and $\eta>0$.

In order to do perturbation theory we have to compute some of the expressions for $\lambda=0$ and $\eta=0$ first.
For an energy $E\in I_{A,K}$ we obtain from \cite[eqs. (3.19) and (4.7)]{KS} the limit as  $\eta\downarrow 0$ of
$\xi_{0,E+i\eta}$ (point wise and in $\Kk_\infty$),  given by
\begin{equation}
 \xi_{0,E}(\bvp_+^\ot, \bvp_-^\ot)=e^{-i\Tr(A_E \bvp_+^\ot - \bar A_E \bvp_-^\ot )},
\end{equation}
where $A_E$ is the matrix
\begin{equation}
 A_E=\tfrac{1}{2K} \left( (E-A)-i\sqrt{K-(E-A)^2}\right)\;.
\end{equation}
Here we identify numbers with multiples of the unit $m\times m$ matrix.
 Note that $E\in I_{A,K}$ is equivalent to $-\sqrt{K} < E-A < \sqrt{K}$ in the sense of matrices,  and for such energies we get
\begin{equation}
 \bth_{0,E}=
-2(\bpart_+ + \bpart_-) \bxi_{0,E} =
\tfrac{2\sqrt{K-(E-A)^2}}{K} \,\bxi_{0,E}\;.
\end{equation}
Thus, we get
\begin{align}
 \llangle \bxi_{0,E}\,|\,\bxi_{0,E}\rrangle_{0,E} &=
\E\left(\Tr(|G_0(0,0;E)|^2)\right) \\ \notag &=
\Tr \left( 4K\left[(K+1)^2 - 4(E-A)^2\right]^{-1} \right) >
\frac{4mK}{(K-1)^2}\;,
\end{align}
\begin{equation} \label{eq-xi-th}
 \llangle \bxi_{0,E}\, |\, \bth_{0,E}\rrangle_{0,E} =\Tr\left( \left[8\sqrt{ K- (E-A)^2}\right]
\left[(K+1)^2 - 4 (E-A)^2\right]^{-1}\right)>0,
\end{equation}
and
\begin{equation} \label{eq-th-th}
\langle \bth_{0,E}\, |\, \bth_{0,E}\rangle_{0,E} =\Tr\left(\left[16( K- (E-A)^2)\right]
\left[{K((K+1)^2 - 4 (E-A)^2)}\right]^{-1}\right) >0\;.
\end{equation}

We can finally prove Theorem~\ref{theo-main1}.

\begin{proof}[Proof of Theorem~\ref{theo-main1}] We only need to prove Equation \eqref{eq-eta3}.
Recall that the maps
$(\lb, E,\eta) \to \xi_{\lb,E+i\eta} \in {\Kk}_\infty$,
$(\lb,E,\eta)\mapsto\bxi_{\lb,E+i\eta}\in\KK$ and
$(\lb,E,\eta)\mapsto\bth_{\lb,E+i\eta}\in\KK$ are continuous on
$\Xi$, which by construction is an open neighborhood of
$\{(0,E,0)\,:\,E\in I_{A,K}\}$ in $\RR\times\RR\times[0,\infty)$.
Using \eqref{eq-Leibn1} and Dominated Convergence one obtains that the map  
$(\lb,E,\eta) \in \RR\times\RR\times[0,\infty) \mapsto \Bb_{\lb,E+i\eta}\in B(\KK)$ 
is continuous with respect to the strong operator topology.
By Proposition~\ref{prop-K-K} we conclude  that the map   $(\lb,E,\eta) \in \Xi \to
\Bb_{\lb,E+i\eta} M(\xi^{K-1}_{\lb,E+i\eta})\bth_{\lb,E+i\eta}\,\in \KK$ is continuous. 
Thus, it follows from \eqref{eq-<>lz}, the definition of $\llangle \cdot \,|\,\cdot\rrangle_{\lb,z}$,
that the real valued maps (cf. Lemma~\ref{lem-tx-tt})
$(\lb, E,\eta) \to  \llangle \bxi_{\lb,E+i\eta}\, | \,  \bth_{\lb,E+i\eta}\rrangle_{\lb,E+i\eta}$
and $(\lb, E,\eta) \to \llangle \bth_{\lb,E+i\eta}\, | \,  \bth_{\lb,E+i\eta}\rrangle_{\lb,E+i\eta}$
have  continuous extensions to $\Xi$. 
Moreover, by \eqref{eq-xi-th} and \eqref{eq-th-th} these extensions satisfy  
\begin{equation}
\llangle \bxi_{\lb,E+i\eta}\, | \,  \bth_{\lb,E+i\eta}\rrangle_{\lb,E+i\eta}\;>\;0
\qtx{and}
\llangle \bth_{\lb,E+i\eta}\, | \,  \bth_{\lb,E+i\eta}\rrangle_{\lb,E+i\eta}\;>\;0\;, \la{pos5}
\end{equation}
for $(\lb,E,\eta)$ in some open neighborhood of $\{(0,E,0)\,:\,E\in I_{A,K}\}$ in $\RR\times\RR\times[0,\infty)$.
Equation \eqref{eq-eta3} now follows from \eqref{eq-J-eta3}. 
\end{proof}



\end{document}